\documentclass[prd,
,superscriptaddress,
nofootinbib,
tightenlines ]{revtex4}
\usepackage{epsfig}
\usepackage[colorlinks=true,linkcolor=blue,urlcolor=blue,citecolor=blue]{hyperref}
\usepackage{amsmath}
\usepackage{amsfonts}
\usepackage{amssymb}
\usepackage{mathtools}

\newcommand{\ben}{\begin{displaymath}}
\newcommand{\een}{\end{displaymath}}
\newcommand{\bga}{\begin{gathered}}
\newcommand{\ega}{\end{gathered}}
\newcommand{\be}{\begin{equation}}
\newcommand{\ee}{\end{equation}}
\newcommand{\bea}{\begin{eqnarray}}
\newcommand{\eea}{\end{eqnarray}}

\begin{document}
\title{Internal force distributions in 't Hooft-Polyakov monopole and Julia-Zee dyon}
 \author{J.~Yu.~Panteleeva}
  \affiliation{Institut f\"ur Theoretische Physik II, Ruhr-Universit\"at Bochum,  D-44780 Bochum,
 Germany}
\date{\today}
\begin{abstract}
The energy-momentum tensor of the 't Hooft-Polyakov monopole and the Julia-Zee dyon are studied. This tensor contains important information about the pressure and the shear force distributions which define mechanical properties of systems. Obtaining the violation of the local stability criterion for the magnetic monopole and dyon we decompose the EMTs into the long- and short-range parts.  This decomposition depends on the abelian field strength tensor which can not be uniquely defined. We suggest to use the modified 't Hooft definition for the tensor.    
Finally, the long- and short-range parts of the EMTs are computed and new equilibrium equations are obtained. Numerical values for masses, $D$-terms and various mean square radii for the monopole and the dyon are also computed.    

\end{abstract}

\maketitle
\begin{center}
\textit{This paper is dedicated to the bright memory of Maxim Polyakov.}
\end{center}
\section{Introduction}
Magnetic monopole is a hypothetical particle carrying a magnetic charge like an electron carrying  electric charge.  Maxwell formulated his elegant  equations in 1865 without monopole because there was no evidence for it. Despite the fact that the monopole is still experimentally not  detected there is also no proof of its nonexistence. In year 1974  A. Polyakov and G.~'t~Hooft have shown  in Refs.~\cite{Polyakov:1974ek, tHooft:1974kcl} that magnetic monopole exists in all Grand Unified Theories of elementary particles as a static soliton solution of the classical equations of motion.  In year 1975  A. Zee and B. Julia generalised the 't Hooft-Polyakov monopole by including the electric charge, they called such particle a dyon in Ref.~\cite{Julia:1975ff}.

Usually the mass and  related properties of a monopole are studied. In this work we are more interested in mechanical properties of a monopole and a dyon such as pressure, share force distributions and the mechanical stability condition which can be obtained from the corresponding energy-momentum tensor (EMT). Although the  main purpose of the paper is  to study the mechanical  properties of the monopole  and dyon, we are also interested in studying the local stability criterium of an arbitrary system with finite energy obtained in 2016  in Ref.~\cite{Perevalova:2016dln}. The force distributions obtained from the EMT could be useful in experimental search of monopoles and studying the  monopole stability can be helpful  for theoretical  insights into stability of classical solutions in gauge theories.      

The organisation of the paper is as follows. In Sec.~\ref{stabilty} we introduce main definitions corresponding to an EMT and discuss the stability conditions for static and non-static EMT. In Sec.~\ref{monopol} we discuss the EMT of the 't Hooft-Polyakov monopole and its consequences.  We also separate the EMT into the long- and short-range parts in this section introducing the abelian strength tensor. In Sec.~\ref{dyon} we discuss the properties of Julia-Zee dyon. We conclude in the last section. 

\section{Stability conditions}
\label{stabilty}
 \subsection{EMT densities} 
The energy-momentum tensor $T_{\mu\nu}(x)$ can be defined as a functional variation of the matter part of the action in curved space-time with respect to metric $g^{\mu\nu}(x)$ as follows 
 \begin{equation}\label{EMTm}
 T_{\mu\nu}(x)=\left.\dfrac{2}{\sqrt{-g}}\dfrac{\delta S}{\delta g^{\mu\nu}(x)} \right \vert_{g_{\mu\nu=\eta_{\mu\nu}}},
 \end{equation}
 where $\eta_{\mu\nu}$ is Minkowski tensor with the metric signature $(1,-1,-1,-1)$.

The EMT encodes  fundamental properties of a system. The $T_{00}$ component of the EMT defines the energy density in the studied system corresponding to the mass distribution in the static approximation. Integral of  the $T_{00}(r)$ over the full space defines the full energy of a system or  its mass in the rest frame of the system
\be\label{masssystem}
M=\int d^3r T_{00}(r).
\ee
The most interesting part of the EMT for our purpose is the $ij$-components, which define the stress tensor. The $T_{ij}$ components can be associated, according to Ref.~\cite{Polyakov:2002yz}, with  the distribution of the shear forces $s(r)$ and the elastic pressure $p(r)$ inside the system. For a spherically symmetric system in static approximation in three dimensions the stress tensor is decomposed  in $s(r)$ and $p(r)$ as follows
\begin{equation}\label{Decomposition}
T_{ij}(\vec{r})=\left(\frac{r_ir_j}{r^2}-\dfrac{1}{3}\delta_{ij}\right)s(r)+\delta_{ij}p(r).
\end{equation}

The pressure and shear force distributions can be connected with the gravitational $D(t)$-form factor, which is involved in the parametrisation of the matrix element of the EMT operator, see e.g.\ Ref.~\cite{Polyakov:2002yz}. As it follows from Ref.~\cite{Polyakov:2002yz}, the $D$-term (or the Druck-term) $D\equiv D(0)$ can be obtained as 
\be\label{Dterm1}
D=-\dfrac{4M}{15}\int d^3r r^2 s(r)=M\int d^3r r^2 p(r),
\ee
where $M$ is the mass of the system.
The value of the $D$-term for a particle can be  measured  experimentally, like its mass, however, the value is difficult to extract from experimental data.  For example, the experimental value of the $D$-term for the proton is $D=-1.47\pm 0.06\pm 0.14$, where the first error is the statistical  uncertainty, and the second error is due to the systematic uncertainties, see Ref.~\cite{Burkert:2021ith}. The value of the $D$-term for the nucleon was first obtained theoretically in the Skyrme model in Ref.~\cite{Cebulla:2007ei}, the authors got  the value $D=-3.6$ and in the bag model the value of the $D$-term for the nucleon is $D=-1.1$ \cite{Hudson:2017oul}, other values of the $D$-term together with the corresponding references can be found in the table 2 of Ref.~\cite{Dutrieux:2021nlz}. The values of the $D$-term for the monopole and the dyon are given in Secs.~\ref{monopol} and \ref{dyon}, respectively. 

 \subsection{Static EMT}
 The EMT is the conserved Noether current associated with spacetime translations,  i.e. it satisfies the following condition
 \be\label{ConservT}
 \partial_\mu T^{\mu\nu}=0.
 \ee
 For the static EMT the Eq.~\eqref{ConservT} turns to 
 $\partial_iT^{i\nu}=0$.
 For the parametrisation of Eq.~\eqref{Decomposition} it implies that the pressure and shear force distributions  satisfy the following equation which is also called equilibrium equation 
 \be \label{StabilityEq}
 \dfrac{\partial p(r)}{\partial r}+\dfrac{2}{3}\dfrac{\partial s(r)}{\partial r}+\dfrac{2}{r}s(r)=0.
 \ee
Assuming that the pressure and shear force densities decay at large distances faster than $\sim\dfrac{1}{r^3}$, multiplying on $r^3$ and integrating over $r$ one obtains the von Laue stability condition given in Ref.~\cite{Laue}
 \be \label{Laucond}
 \int d^3r p(r)=0,
 \ee
 which is necessary  for stability, but not sufficient, since it is also satisfied for unstable  systems,  see discussion in Ref.~\cite{Perevalova:2016dln}. This condition is satisfied for any system, whose EMT is conserved and implies that the pressure distribution has at least one mode. 
Integrating the differential equation \eqref{StabilityEq}, from positive value $r$ to infinity and requiring that $s(r)$ and $p(r)$ vanish at the infinity, one obtains
 \be\label{NormForce}
\dfrac{2}{3}s(r)+p(r)=2\int\limits_r^{\infty}dx\dfrac{s(x)}{x}.
 \ee
 This equation describes the equilibrium of the internal forces inside a system.  
 The combination $ \frac{2}{3}s(r)+p(r)$ describes the normal component of the total force exhibited by the system on an infinitesimal piece of area $dS_i$, which is denoted as $p_{\text{r}}(r)$:
 \be
 F^i(r)=T^{ij}(r)dS_j=\left(\frac{2}{3}s(r)+p(r)\right)dS^i=p_{\text{r}}(r)dS^i
 \ee
where $dS^i=dS\ r^i/r$. From the equilibrium equation \eqref{NormForce} follows that for the positive shear force distribution $s(r)$ the normal force is always positive, for the negative $s(r)$ the normal force is always negative. In Ref.~\cite{Perevalova:2016dln} it was argued that the normal force has to be directed outwards otherwise the system would collapse, this means that it has to be non-negative, i.e.  
 \be\label{Mstability}
 \dfrac{2}{3}s(r)+p(r)\geq 0.
 \ee
Thus, according to the Eq.~\eqref{NormForce} 
  for negative shear force distribution the system is definitely unstable. The local stability  condition is also necessary, but not sufficient for stability analogously to the von Laue condition. However, due to its local character, it is stronger than the von Laue condition.

 \subsection{EMT and external forces}
The EMT conservation taking the form $\partial_iT^{ij}(r)=f^j(r)$
can be interpreted according to Ref.~\cite{Landau} as the equilibrium equation for internal stress and external force  $f_j$ (per unit of the volume). According to the parametrisation in Eq.~\eqref{Decomposition} the equilibrium equation gets the following form 
\be\label{equilibriumStars}
\dfrac{d}{dr}\left(\dfrac{2}{3}s(r)+p(r)\right)+2\dfrac{s(r)}{r}=f(r),
\ee
where $f(r)$ is the normal component of the external force per unit of the volume. This equation describes the balance between internal forces pushing out from center and  external force pulling inwards to the center. When the corresponding forces are equal, the system is at equilibrium.
 The von Laue stability condition  for such equilibrium equation gets the following form
 \be
 \int d^3r p(r)=- \dfrac{1}{3}\int d^3r~rf(r).
 \ee 
One can also rewrite the equilibrium equation \eqref{equilibriumStars} as
 \be
p_{\text{r}}(r)+\sigma(r)=2\int\limits_r^{\infty}dx\dfrac{s(x)}{x},
 \ee 
with $\sigma(r)=\int\limits_r^{\infty}dx f(x)$. The left-hand side describes the normal component of a total force of the system acting on the infinitesimal unit area $dS^i$ as it follows from below
\be
F^i(r)=(T^{ij}(r)+\delta^{ij}\sigma(r))dS_j=\left(\frac{2}{3}s(r)+p(r)+\int\limits_r^{\infty}dx f(x)\right)dS^i=\Big(p_{\text{r}}(r)+\sigma(r)\Big)dS^i.
\ee
Then for systems affected by an external force, i.e. for systems described by the equilibrium equation \eqref{equilibriumStars},  the local stability criterium of Eq.~\eqref{Mstability} can be modified as follows  
 \be\label{stabilityJ}
\dfrac{2}{3}s(r)+p(r)+\int\limits_r^{\infty}dx f(x)\geq 0.
 \ee 
For a time-dependent system the conservation of EMT gives the following equation $\partial_i T^{ij}=-\partial_0 T^{0j}$, where the left-hand side again describes internal  force of the system and the right-hand side can be interpreted as an external force.

 \subsection{Mean square energy radius  and mechanical radius}
For a positive energy density $T_{00}(r)$ the mean square radius of the energy density can be  introduced as
\be\label{r^2E}
\langle r^2\rangle_E=\dfrac{\int d^3r r^2T_{00}(r)}{\int d^3r T_{00}(r)},
\ee   
which characterises the size of the system in which the energy is distributed.

According to the local stability condition, for a stable system the radial force must be positive, so  the mechanical mean square radius where the normal force is distributed  can be defined as in Ref.~\cite{Polyakov:2018zvc}, for a system with the equilibrium equation \eqref{StabilityEq} it takes the following form 
\be\label{r^2mech}
\langle r^2\rangle_{\text{mech}}=\dfrac{\int d^3r\ r^2 p_\text{r}(r) }{\int d^3r p_\text{r}(r)},
\ee 
and for the system with the equilibrium equation \eqref{equilibriumStars} it takes has the form:
\be\label{r^2mech2}
\langle r^2\rangle_{\text{mech}}=\dfrac{\int d^3r\ r^2 \left(p_\text{r}(r)+\sigma(r)\right) }{\int d^3r \left(p_\text{r}(r)+\sigma(r)\right)},
\ee 
where $p_\text{r}(r)$ and $\sigma(r)$ are defined above.

The local stability condition has not been mathematically proven and thereby is still questioned, see e.g. criticism in Ref.~\cite{Ji:2021mfb}. Moreover, as it was mentioned in recent studies of Refs.~\cite{Varma:2020crx, Varma:2022kbv, Loiko:2022noq}, the local stability condition is not satisfied in the presence of the long-range forces. Additional to that, as pointed out in Refs.~\cite{Varma:2020crx, Varma:2022kbv, Metz:2020vxd, Freese:2022jlu}, there is another  problem for systems with the long-range contribution, namely, the divergence of essential quantities, that  describe mechanical properties of a system, such as the $D$-term and  the mean square radii corresponding to the EMT densities.

At this point, it is important to mention, that the 't Hooft-Polyakov monopole is accepted to be a stable system, see arguments of Refs.~\cite{Baacke:1990at,Gervalle:2022npx}. As we will see later, the local stability condition is violated for the monopole and the dyon. On the first glance, thereby one could think that the criterium is not correct. However,  the monopole and the dyon involve electromagnetic interaction, which supports the idea that the local stability condition does not apply correctly in the presence of long-range forces. 
 
\section{'t Hooft-Polyakov monopole}
\label{monopol}
\subsection{Equations of motion}
\label{action2}
 The Grand Unified Theories combine the electromagnetic, weak, and strong forces into a single force. One of the first such theories  was suggested by   Howard Georgi and Sheldon Glashow in 1974 in Ref.~\cite{Georgi:1974sy}. Later  Alexander Polyakov and Gerard 't Hooft have independently found that magnetic  monopoles automatically appear in all Grand Unified Theories \cite{Polyakov:1974ek,tHooft:1974kcl}. The most simple model, where  magnetic monopole exists is  the  gauge SU(2) Georgi-Glashow model with  Higgs triplet field $\varphi^a,\  a=1,2,3$, which belongs to adjoint representation.
The corresponding gauge invariant action of the model is
\begin{equation}\label{action}
S=\int d^4x\left[-\dfrac{1}{4} F_{\mu\nu}^aF^{\mu\nu a}+\dfrac{1}{2}(\mathcal{D}_{\mu}\varphi)^a(\mathcal{D}^{\mu}\varphi)^a-\dfrac{\lambda}{4}(\varphi^a\varphi^a-v^2)^2\right],
\end{equation} 
where $\lambda$ is a dimensionless coupling constant, $v^2$ is the squared vacuum expectation value  of the Higgs field; $\mu$ and $\nu$ are Lorenz indices. 
The covariant derivative and the non-abelian field  strength  tensor are defined as follows
\begin{equation}\label{derive}
\begin{split}
 (\mathcal{D}_{\mu}\varphi)^a=\partial_{\mu}\varphi^a+g\epsilon^{abc}A^b_{\mu}\varphi^c,\\
 F_{\mu\nu}^a=\partial_{\mu}A_{\nu}^a-\partial_{\nu}A_{\mu}^a+g\epsilon^{abc}A_{\mu}^bA_{\nu}^c,
 \end{split}
 \end{equation}
where $g$ is the gauge coupling constant and  $A_{\mu}^a$ is the gauge vector  field.
Choosing the vacuum expectation value as $\varphi_0^a=\left(0,0,v\right)$ and considering small fluctuations of the ground state in unitary gauge, one can show  that  the Georgi-Glashow model has one massive scalar field $\eta_3(x)$ with mass $m_H=\sqrt{2\lambda}v$, one massless vector field $A_{\mu}^3$ corresponding to the $U(1)$ subgroup of $SU(2)$ gauge group and two massive vector fields $A_{\mu}^1$ and $A_{\mu}^2$  both with masses  $m_V=gv$.

In this model we are interested in a soliton solution, in other words, in a static solution of classical field equations with finite energy.  As we want to study static soliton configuration, we consider the fields $A_i^a(\vec{x})$ and $\varphi^a(\vec{x})$  independent of time.  We also fix zero component of a vector field through the gauge $A_0^a=0$ to have zero electric field. 
Requiring the energy to be finite the following configuration of fields can be found
\begin{equation}\label{anzatz}
\begin{split}
\varphi^a&=n^av h(r),\\
A^a_i&=\dfrac{1}{gr}\epsilon_{aij}n_j(1-F(r)),
\end{split}
\end{equation}
with the unit vector $n^a=\frac{r^a}{r}$, the unknown profile functions $h(r)$ and $F(r)$ and the following boundary conditions
\begin{equation}\label{boundcon}
\bga
F(r)=0\hspace{10pt}\text{at}\hspace{10pt}r\rightarrow\infty,\hspace{20pt} F(r)=1\hspace{10pt}\text{at}\hspace{10pt}r\rightarrow 0,\\
h(r)=1\hspace{10pt}\text{at}\hspace{10pt}r\rightarrow\infty,\hspace{20pt} h(r)=0\hspace{10pt}\text{at}\hspace{10pt}r\rightarrow 0.
\ega
\end{equation}
The explicit form of the profile functions can be obtained from equations of motion which can be computed  by varying the action  with respect to scalar and vector fields, which  for the static case reduce to the following form
\begin{equation}
\begin{split}
\mathcal{D}_{i}F_{ij}^a&=g\epsilon^{abc}\varphi^b(\mathcal{D}_{j}\varphi)^c,\\
\mathcal{D}_{i}(\mathcal{D}_{i}\varphi)^a&=\lambda(\varphi^b\varphi^b-v^2)\varphi^a.
\end{split}
\end{equation} 
In terms of the profile functions the equations transform to
\begin{equation}\label{eq1}
\begin{split}
F''(r)-\dfrac{F(r)(F^2(r)-1)}{r^2}-g^2 v^2 F(r)h^2(r)&=0,\\
h''(r)+2\dfrac{h'(r)}{r}-2\dfrac{F^2(r)h(r)}{r^2}+\lambda v^2 h(r)\left(1-h^2(r)\right)&=0.
\end{split}
\end{equation}
Rescaling the argument $r$ with dimensionless argument $\rho$ as
$\rho=\frac{r}{R_0}$, where $R_0=\frac{1}{m_V}=\frac{1}{gv}$  is a typical size of the solution, and introducing the new parameter 
$\beta^2=2\frac{\lambda}{g^2}=\frac{m_H^2}{m_V^2}$
we obtain  the following system of equations of motion
\begin{equation}\label{fulleqs}
\begin{split}
F''(\rho)-\dfrac{F(\rho)(F^2(\rho)-1)}{\rho^2}- F(\rho)h^2(\rho)&=0,\\
h''(\rho)+2\dfrac{h'(\rho)}{\rho}-2\dfrac{F^2(\rho)h(\rho)}{\rho^2}+\dfrac{\beta^2}{2}h(\rho)\left(1-h^2(\rho)\right)&=0,\\
\end{split}
\end{equation} 
with the same boundary conditions as before. This system can be solved analytically only for the limit $\beta=0$, otherwise one solves it numerically. However, one can find approximate solutions at small and large distances: 
\begin{equation}
\label{sollim2}
\begin{split}
F(\rho)&\underset{\rho\to\infty} \simeq C_Fe^{-\rho}\left(1-\dfrac{1}{2\rho}+\dfrac{3}{8\rho^2}+O\left(\dfrac{1}{\rho^3}\right)\right),\hspace{8pt} F(\rho)\underset{\rho\to 0}\simeq 1+a \rho^2+\sum_{n=2}^{\infty} c_n \rho^{2n},\\
h(\rho)&\underset{\rho\to\infty} \simeq 1-C_h\dfrac{e^{-\beta\rho}}{\rho}\left(1+O\left(\dfrac{1}{\rho}\right)\right)-\dfrac{2C_F^2}{\beta^2-4}\dfrac{e^{-2\rho}}{\rho^2}\left(1-O\left(\dfrac{1}{\rho}\right)\right),\hspace{8pt} h(\rho)\underset{\rho\to 0}\simeq b \rho+\sum_{n=2}^{\infty}d_n \rho^{2n-1},
\end{split}
\end{equation}
where $C_F$, $C_h$, $a$ and $b$ are free constants. Note, that for $\beta=2$
 the constant $C_F=0$. The coefficients $c_n$ and $d_n$ can be expressed in terms of  $a$ and $b$ as follows
\begin{equation}
\begin{split}
c_2=&\dfrac{1}{10}\left(3a^2+b^2\right),\hspace{5pt} c_3=\dfrac{1}{70}\left(7 a^3 + 6 a b^2-\dfrac{b^2\beta^2}{4}\right),\\
d_2=&\dfrac{b}{10}\left(4a-\dfrac{\beta^2}{2}\right),\hspace{5pt} d_3=\dfrac{1}{280} \left(48 a^2 b + 4 b^3 - 4 a b\beta^2+ 5 b^3 \beta^2+ b \dfrac{\beta^2}{2}\right).
\end{split}
\end{equation}
In Ref.~\cite{Korenblit:2021nzq} another system of differential equations is obtained and analytically solved for complex non-abelian monopole and dyon fields. 

\subsection{EMT densities}
The EMT for the 't Hooft-Polyakov monopole can be obtained variating generally covariant form of Georgi-Glashow action \eqref{action} with respect to the metric
\begin{equation}
T_{\mu\nu}=-\eta_{\mu\nu}\mathcal{L}+F_{\mu\alpha}^aF^{\alpha a}_{\ \nu}+(\mathcal{D}_{\mu}\varphi)^a(\mathcal{D}_{\nu}\varphi)^a.
\end{equation}
For the static case one gets 
\begin{equation}\label{T00}
\begin{split}
T_{00}&=\dfrac{1}{4}F_{ij}^aF_{ij}^{ a}+\dfrac{1}{2}(\mathcal{D}_i\varphi)^a(\mathcal{D}_i\varphi)^a+\dfrac{\lambda}{4}(\varphi^a\varphi^a-v^2)^2,\\
T_{ij}&=-\dfrac{1}{4}\delta_{ij}F_{km}^aF_{km}^{ a}-F_{ik}^aF_{kj}^{ a}-\dfrac{1}{2}\delta_{ij}(\mathcal{D}_{k}\varphi)^a(\mathcal{D}_{k}\varphi)^a+(\mathcal{D}_i\varphi)^a(\mathcal{D}_j\varphi)^a-\dfrac{\lambda}{4}\delta_{ij}(\varphi^a\varphi^a-v^2)^2.
\end{split}
\end{equation}
Using the 't Hooft-Polyakov ansatz given in Eq.~\eqref{anzatz}  and the decomposition in Eq.~\eqref{Decomposition} the EMT densities can be expressed in terms of the profile functions
$F(\rho)$ and $h(\rho)$:  
\begin{equation}\label{energy}
\begin{split}
T_{00}(\rho)&=\dfrac{1}{R_0^4g^2}\left(\dfrac{F'^2}{\rho^2}+\dfrac{(1-F^2)^2}{2\rho^4}+\dfrac{1}{2}h'^2+\dfrac{h^2F^2}{\rho^2} + \dfrac{\beta^2}{8}(1-h^2)^2\right),\\
s(\rho)&=\dfrac{1}{R_0^4g^2}\left(\dfrac{F'^2}{\rho^2}-\dfrac{(1-F^2)^2}{\rho^4}+h'^2-\dfrac{1}{\rho^2}F^2h^2\right),\\
p(\rho)&=\dfrac{1}{R_0^4g^2}\left(\dfrac{1}{3}\dfrac{F'^2}{\rho^2}+\dfrac{(1-F^2)^2}{6\rho^4}-\dfrac{1}{6}h'^2-\dfrac{1}{3}\dfrac{F^2}{\rho^2}h^2-\dfrac{\beta^2}{8}(1-h^2)^2\right).
\end{split}
\end{equation}
The  $T_{00}(\rho)$ component provides the information about spatial distribution of the monopole mass, $p(\rho)$ and $s(\rho)$ describe the pressure and shear force distributions inside the monopole. The first two terms in these densities are originating from the term $\sim F^a_{\mu\sigma}F^{a\sigma}_{\phantom{a\sigma}\nu}$ in action, the next two terms are coming  from $\sim\mathcal{D_{\mu}}\varphi^a\mathcal{D_{\nu}}\varphi^a$, and the term proportional to $\beta^2$ comes from the Higgs potential.

Before we discuss the numerical results for the EMT density distributions and their properties, we compute their asymptotic behaviour at large and small distances using the behaviour of the profile functions from Eq.~\eqref{sollim2}:  
\begin{equation}\label{EMTinf}
\begin{split}
T_{00}(\rho)&\underset{\rho\to\infty}\simeq \dfrac{1}{R_0^4g^2}\left(\dfrac{1}{2\rho^4}+\beta^2 C_h^2 \dfrac{e^{-2\beta \rho}}{\rho^2}\left[1+O\left(\dfrac{1}{\rho}\right)\right]+2C_F^2\dfrac{e^{-2\rho}}{\rho^2}\left[1+O\left(\dfrac{1}{\rho}\right)\right]\right),\\
s(\rho)&\underset{\rho\to\infty}\simeq \dfrac{1}{R_0^4g^2 }\left(-\dfrac{1}{\rho^4}+\beta^2 C_h^2 \dfrac{e^{-2\beta \rho}}{\rho^2}\left[1+O\left(\dfrac{1}{\rho}\right)\right]+C_F^2\dfrac{e^{-2\rho}}{\rho^4}\left[1+O\left(\dfrac{1}{\rho}\right)\right]\right),\\
p(\rho)&\underset{\rho\to\infty}\simeq  \dfrac{1}{R_0^4g^2}\left(\dfrac{1}{6\rho^4}-\dfrac{2}{3}\beta^2C_h^2\dfrac{e^{-2\beta\rho}}{\rho^2}\left[1+O\left(\dfrac{1}{\rho}\right)\right]-\dfrac{2C_F^2}{3}\dfrac{e^{-2\rho}}{\rho^4}\left[1+O\left(\dfrac{1}{\rho}\right)\right]\right),
\end{split}
\end{equation}
and
\begin{equation}\label{EMTorigin}
\begin{split}
T_{00}(\rho)&\underset{\rho\to 0}\simeq \dfrac{1}{R_0^4g^2}\left(\left[6a^2+\dfrac{3 b^2}{2}+\dfrac{\beta^2}{8}\right] +\rho^2\left[ 8a^3+6ab^2-\dfrac{\beta^2b^2}{2}\right]+O(\rho^4)\right),\\
s(\rho)& \underset{\rho\to 0}\simeq \dfrac{1}{R_0^4g^2}\left(\left[-8a^3+2ab^2-\beta^2b^2\right]\dfrac{\rho^2}{5}+O(\rho^4)\right),\\
p(\rho)&\underset{\rho\to 0}\simeq  \dfrac{1}{R_0^4g^2}\left( \left[2a^2-\dfrac{b^2}{2}-\dfrac{\beta^2}{8}\right] +\dfrac{\rho^2}{3}\left[8a^3-2ab^2+\beta^2b^2\right]+O(\rho^4)\right),
\end{split}
\end{equation}
where $C_h$, $C_F$ and  $a$, $b$ are free constants from the asymptotic behaviour of profile functions in Eq.~\eqref{sollim2}. 
From the large distances behaviour it is clear that the power-law decay ($\sim \rho^{-4}$) is dominating, this behaviour corresponds to the contribution of the electromagnetic interaction.\footnote{We will call this contribution the long-range interaction.} This is not surprising if one remembers that  the mass spectrum of the Lagrangian \eqref{action} has one massless vector particle corresponding to the $U(1)$ group. 

The asymptotic behaviour of the energy density is definitely positive at the infinity as well as at the origin. This hints that we are dealing with an usual system. The outer region of the pressure distribution has definitely positive sign  and that of the shear force distribution -- negative one. Such behaviour leads to violation of the local stability criterion \eqref{Mstability}.  
However, this behaviour is related only to the electromagnetic contribution. Note, that  because of this long-range contribution both mean square radii in Eqs.~\eqref{r^2E} and  \eqref{r^2mech} as well as the Druck-term in Eq.~\eqref{Dterm1} diverge. 

\begin{figure}[ht]
\begin{minipage}[h]{0.495\linewidth}
\center{\includegraphics[width=0.81\linewidth]{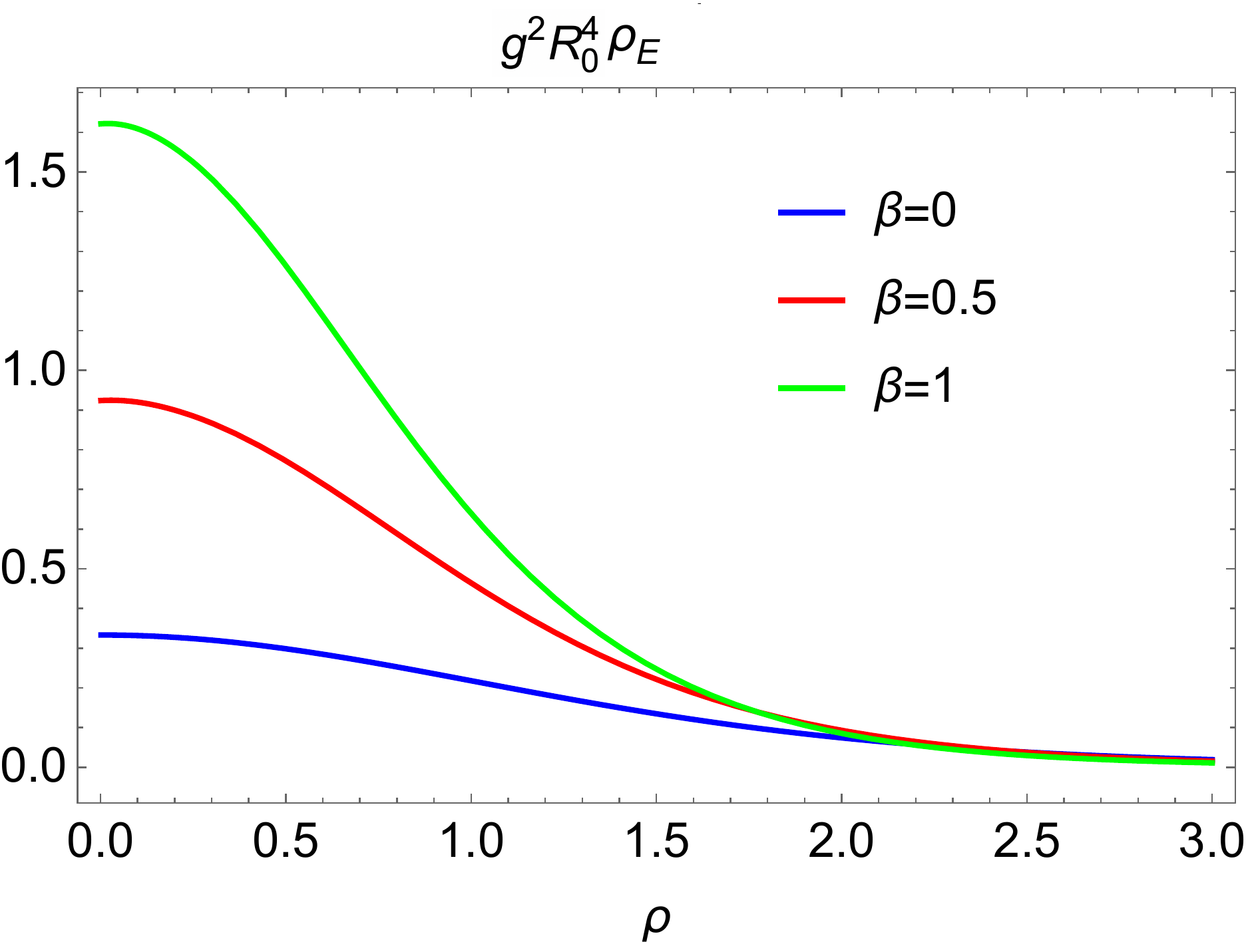}} \\a.
\end{minipage}
\hfill
\begin{minipage}[h]{0.495\linewidth}
\center{\includegraphics[width=0.85\linewidth]{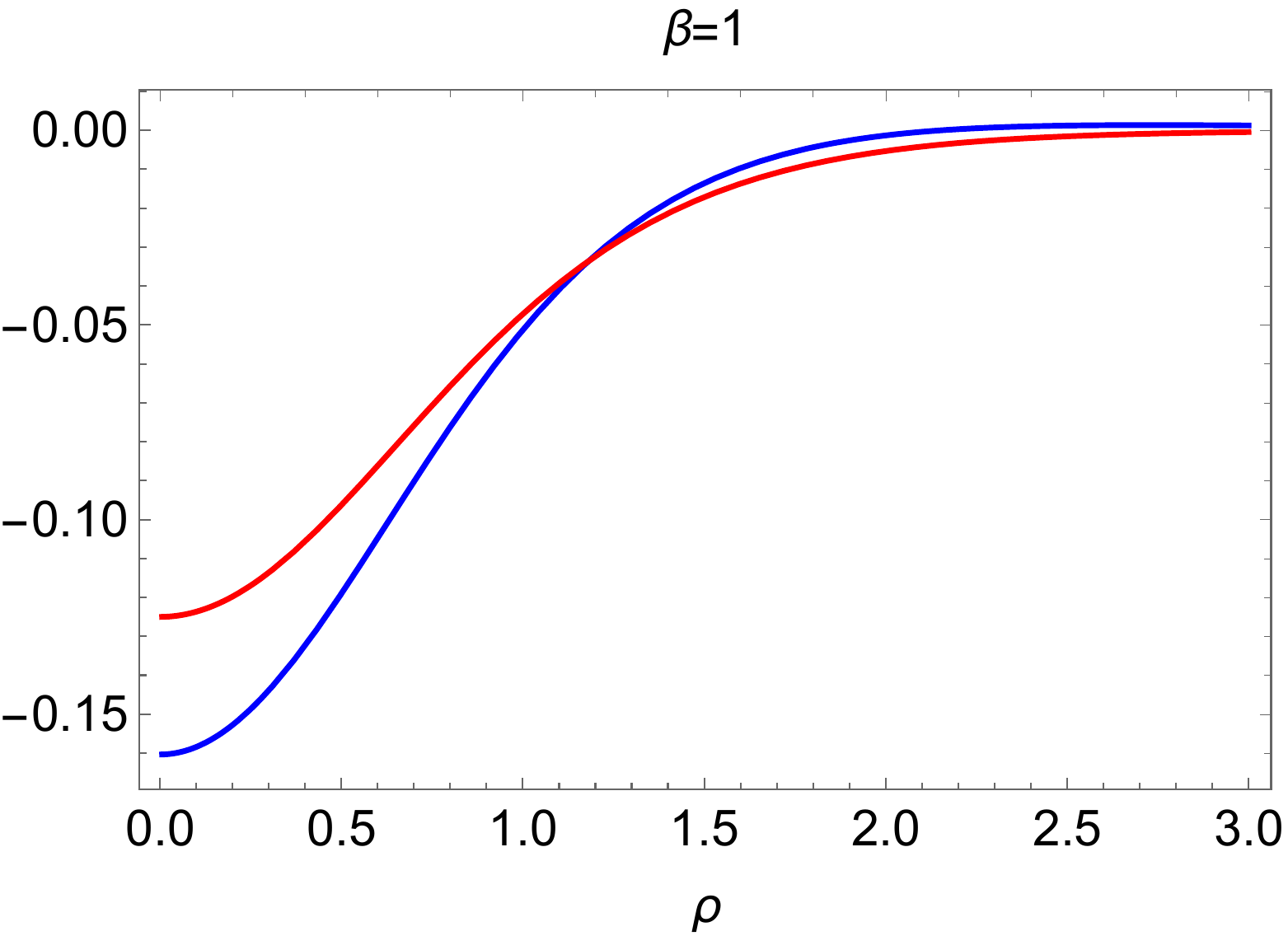}} \\b.
\end{minipage}
\caption{a. Energy distribution  as a function of $\rho=r/R_0=gvr$ for various values of  $\beta$. \\
b. The blue line: the full pressure distribution $p(\rho)$ as a function of $\rho=r/R_0=gvr$ and the red line: the contribution of the Higgs potential only to the full pressure.}
\label{pdistrib}
\end{figure}

\begin{figure}[ht]
\begin{minipage}[h]{0.495\linewidth}
\center{\includegraphics[width=0.87\linewidth]{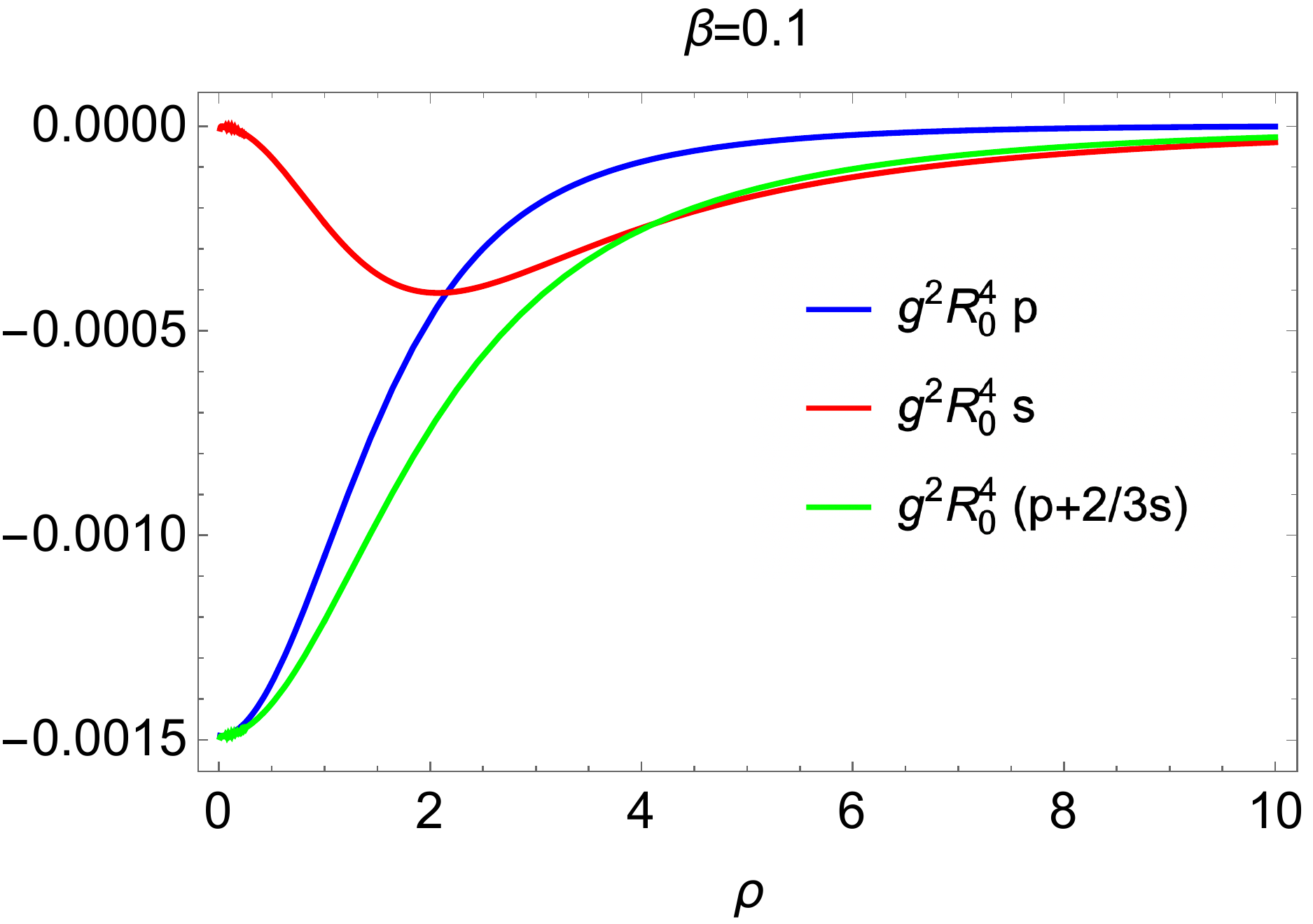}} 
\end{minipage}
\hfill
\begin{minipage}[h]{0.495\linewidth}
\center{\includegraphics[width=0.81\linewidth]{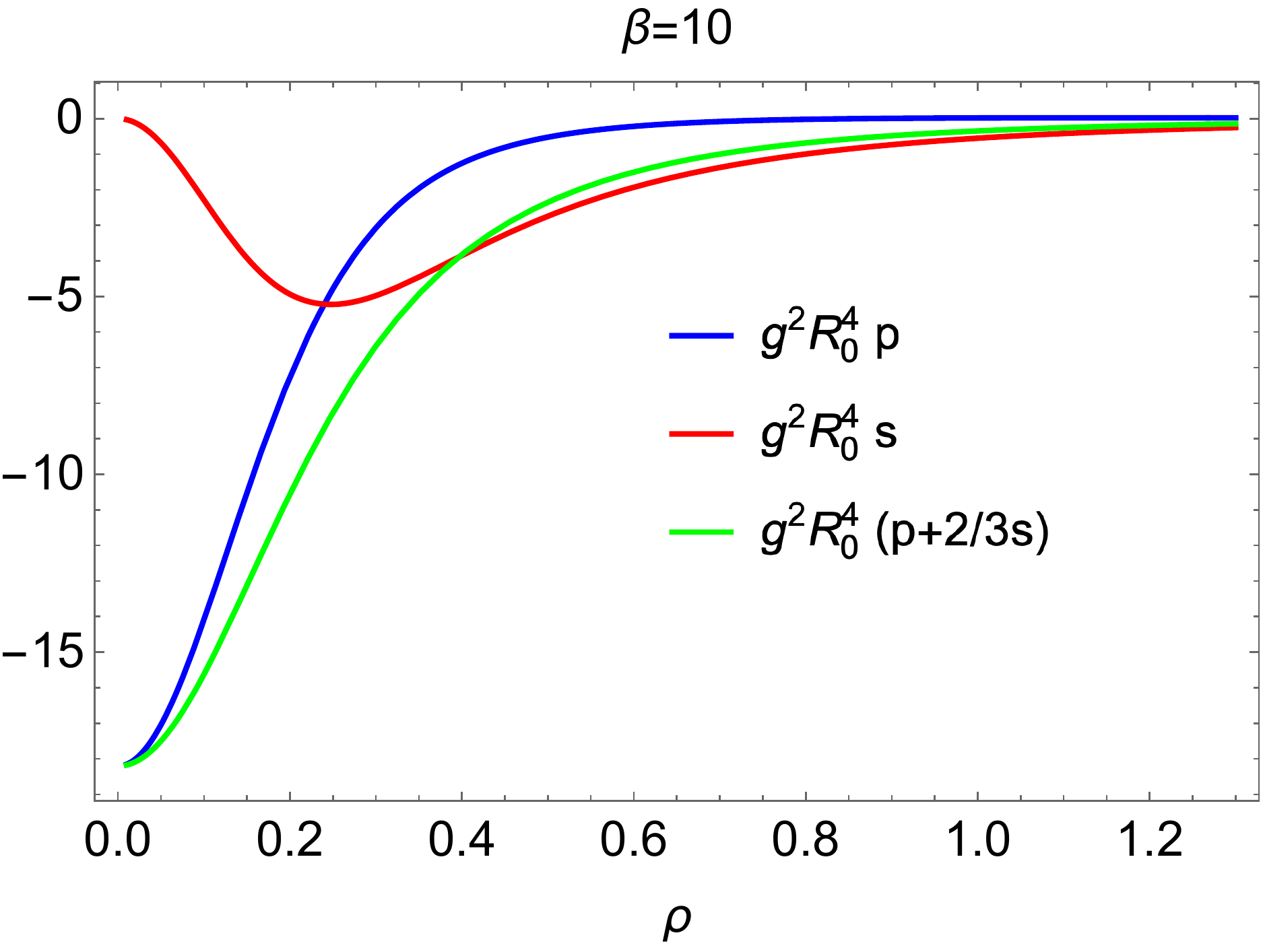}} 
\end{minipage}
\caption{The pressure distribution, the shear force distribution and stability condition from Eq.~\eqref{Mstability} as functions of $\rho=r/R_0=gvr$ for various values of $\beta$.}
\label{oldemt}
\end{figure}

In Fig.~\ref{pdistrib} the energy density  for various values of $\beta$ is shown.  Although the energy is growing with $\beta$, according to Ref.~\cite{Kirkman:1981ck} it does not diverge for $\beta\rightarrow\infty$.
The distributions of the shear force, pressure and normal force are shown in Fig.~\ref{oldemt}. Although the pressure is mostly negative, it changes the sign which  allows the von Laue condition  \eqref{Laucond} to be satisfied, see e.g. Fig.~\ref{PressureNul}. It is also interesting  that the maximum of pressure is three times  larger than the maximum of  shear forces like in many other systems studied before, see e.g Refs.~\cite{Varma:2020crx,Cebulla:2007ei,Perevalova:2016dln,Polyakov:2018zvc,Polyakov:2002yz}. 
From the figure one also sees that the stability condition \eqref{Mstability} is violated everywhere and for every choice of $\beta$, although it is proved in Refs.~\cite{Baacke:1990at,Gervalle:2022npx} that the monopole is stable. Since the local  stability condition is not mathematically proved, we cannot conclude from its violation that the monopole is unstable. This violation can be also explained by the fact that the stability criterion  is inapplicable for systems with  the long-range force that is present in the monopole. 
The authors of Refs.~\cite{Varma:2020crx, Varma:2022kbv} also obtained the violation of the stability criterion for stable systems like proton in the presence of long-range forces. All these aspects motivate us to think that the stability condition in Eq.~\eqref{Mstability} can be applied only for the systems where only the short-range interactions are present. 

\subsection{BPS limit}
\label{BPSLim}
The equations of motion can be solved analytically only  for the limit $\beta=0$. The solution of these equations was first obtained  in Ref.~\cite{Prasad:1975kr}. In Ref.~\cite{Bogomolny} Bogomolny  solved these equations by reducing the system of the second order equations to the first order by considering the energy functional for the Georgi-Glashow model.  By integrating the energy density over the volume one gets the energy functional
 \be
E_{\beta=0}=\int d^3rT_{00}^{\beta=0}(r)=\int d^3r\left[\dfrac{1}{4} F_{ij}^aF_{ij}^a+\dfrac{1}{2}(\mathcal{D}_{i}\varphi)^a(\mathcal{D}_{i}\varphi)^a\right].
\ee
Rewriting it in terms of the chromomagnetic field $H_i^a=-\frac{1}{2}\epsilon_{ijk}F^a_{jk}$,
Bogomolny introduced  the following inequality for the monopole energy, which is also called BPS bound 
\be\label{BSPEng}
E_{\beta=0}\geq\int d^3r H_i^a\mathcal{D}_i\varphi^a,
\ee
where the Bianchi identity $\left(\mathcal{D}_iH_i\right)^a=0$ and the Gauss's theorem were used. The BPS  limit gives the possible minimum of the static energy of the monopole and the equality holds for the Bogomolny equation
\be\label{BogEq}
\mathcal{D}_i\varphi^a=H_i^a.
\ee
From the definition of chromomagnetic field the following  expression can be obtained
\be
H_i^aH_j^a=F^a_{ik}F^a_{kj}+\dfrac{1}{2}\delta_{ij}F^a_{km}F^a_{km}.
\ee 
Using this expression and Bogomolny equation \eqref{BogEq} it can be shown that the spatial components of the EMT disappear:
\be
T_{ij}(r)=0,
\ee
independently of the choice of the  ansatz in Eq.~\eqref{anzatz}.
 
The vanishing of the pressure and shear force distributions at the BPS limit allows to assume that the main contribution for the stress tensor $T_{ij}$ is given by the Higgs potential.  As it can be seen from Fig.~\ref{pdistrib} the largest contribution to the pressure distribution is indeed due to the term originating from the Higgs potential in Eq.~\eqref{action}. Although the shear force distribution does not depend explicitly on $\beta$, it is related to the pressure distribution due to the differential equation \eqref{StabilityEq}.
We think that in all models where it is possible to construct the BPS limit, the corresponding $ij$~components of EMT would vanish. We have also obtained such vanishing, for example,  for baby Skyrme model. The vanishing of the pressure and shear force distributions for BSP limit indicates that the 't Hooft-Polyakov monopole in this limit has the isotropic matter distribution~\cite{Polyakov:2018zvc}.
  
\subsection{Abelian field strength tensor and electromagnetic EMT}
In the previous subsection we  obtained the EMT in non-abelian  $SU(2)$ gauge theory, however, the theory has one massless vector field  corresponding to the abelian $U(1)$ subgroup. This massless vector field is responsible for the electromagnetic contribution to the EMT and  for the divergence of such mechanical properties of the monopole as the $D$-term, mean square energy and mechanical radii.  We will subtract this long-range contribution from the EMT and study the remaining short-range structure. For this we have to determine the $U(1)$ abelian field strength tensor. 
Let us denote the abelian field strength tensor as $\mathcal{F}_{\mu\nu}$. Since in the unitary gauge the massless vector field of the theory is the third component of the vector field $A_3^{\mu}$ (see Sec.\ref{action2}), the expression for the abelian field strength tensor must be $SU(2)$ gauge invariant and coincide with $\mathcal{F}_{\mu\nu}=\partial_\mu A^3_\nu-\partial_\nu A_\mu^3$ in the unitary gauge~\cite{tHooft:1974kcl}. 
We introduce a general definition of the abelian field strength tensor without any requirements on $\phi$ as
\be
\mathcal{F}_{\mu\nu}=\phi^aF^a_{\mu\nu}-\dfrac{c_1}{g}\varepsilon^{abc}\phi^a\mathcal{D}_{\mu}\phi^b\mathcal{D}_\nu\phi^c.
\ee
The constant $c_1$ is not fixed by the above requirements  for  the abelian field strength tensor  $\mathcal{F}_{\mu\nu}$, however, it is needed to define the magnetic charge density in such a way that it coincides with the topological charge density.
G. 't Hooft suggested in Ref.~\cite{tHooft:1974kcl} the tensor with the unit vector $\phi^a=\hat{\varphi}^a=\frac{\varphi^a}{|\varphi|}$ and $c_1=1$, Faddeev offered  in Ref.~\cite{Faddeev} the tensor with   $\phi^a=\dfrac{\varphi^a}{v}$ and $c_1=0$, where $\varphi^a$ is the ansatz from Eq.~\eqref{anzatz} and Boulware in Ref.~\cite{Boulware:1976tv} suggested the tensor with $\phi^a=\hat{\varphi}^a=\frac{\varphi^a}{|\varphi|}$ and $c_1=0$. Since the under-lying theory combines the long- and short-range forces it is not possible uniquely defined the $U(1)$ field strength tensor to separate the short-range interaction from the long one.

For the general choice of the abelian field strength tensor the magnetic charge density  gets the following form
\be
\bga\label{ChargeDens}
\rho_M=-\dfrac{1}{2}\varepsilon_{ijk}\partial_i\mathcal{F}_{jk}=\\
=\dfrac{1}{2}\varepsilon_{ijp}(\mathcal{D}_p\phi)^aF^a_{ij}(c_1\vec{\phi}^2-1)+\dfrac{c_1}{2g}\varepsilon^{abc}\varepsilon_{ijp}(\mathcal{D}_i\phi)^a(\mathcal{D}_j\phi)^b(\mathcal{D}_p\phi)^c-\dfrac{c_1}{2}\varepsilon_{ijp}\phi^aF^a_{ij}\dfrac{1}{2}\partial_p\vec{\phi}^2.
\ega
\ee
One can show that for the 't Hooft's definition of the abelian field strength tensor the magnetic and topological charge densities coincide and  describe single point-like particle with the magnetic charge $\frac{4\pi}{g}$ at the origin, see Ref.~\cite{tHooft:1974kcl}. Such monopole is called Dirac's monopole. In Ref.~\cite{Dirac:1931kp} Dirac derived the quantisation of the magnetic charge as $Q_M=Q_T/g$, where $Q_M$ is a magnetic charge and $Q_T$ is  a topological charge.

In contrast to the 't Hooft's definition of the abelian field strength tensor the Faddeev's and Boulware's definitions describe the magnetic charge density smoothly without singularities at the origin. The topological and magnetic charge densities are not equal in these cases, but the quantisation of the magnetic charge is satisfied.
It is also interesting to notice that  the magnetic charge density in Eq.~\eqref{ChargeDens} for the Faddeev's and and Boulware's   definitions of the abelian strength tensor coincides with the energy density in BSP limit in Eq.~\eqref{BSPEng} up to some dimensional normalisation factor. For the positive magnetic  charge distribution the mean square radius can be determined  
\be\label{chargeradius}
\langle r^2\rangle_M=\dfrac{\int d^3 r r^2\rho_M(r)}{\int d^3 r \rho_M(r)}.
\ee

\begin{figure}[ht]
\begin{minipage}[h]{0.495\linewidth}
\center{\includegraphics[width=0.87\linewidth]{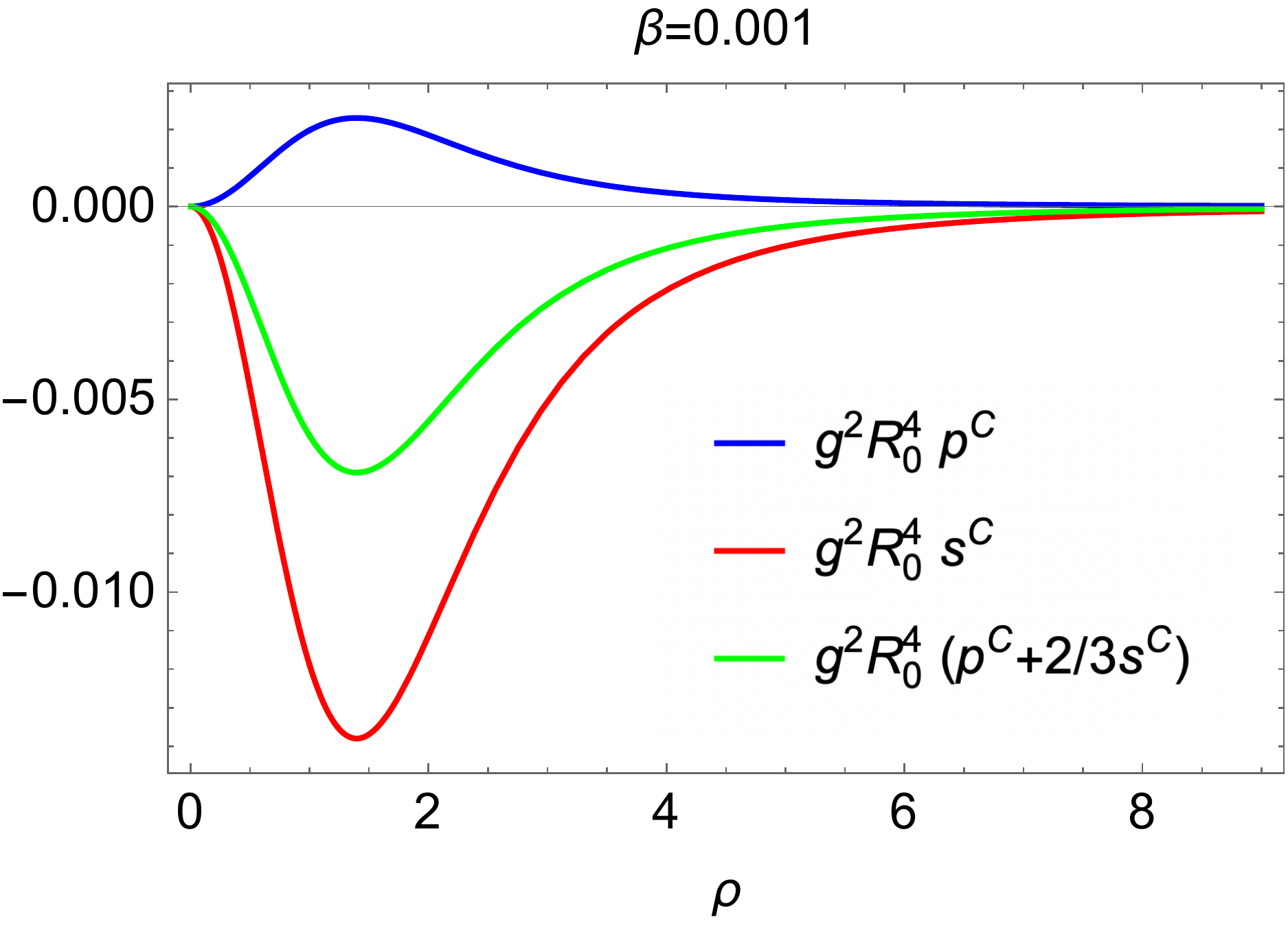}} 
\end{minipage}
\hfill
\begin{minipage}[h]{0.495\linewidth}
\center{\includegraphics[width=0.81\linewidth]{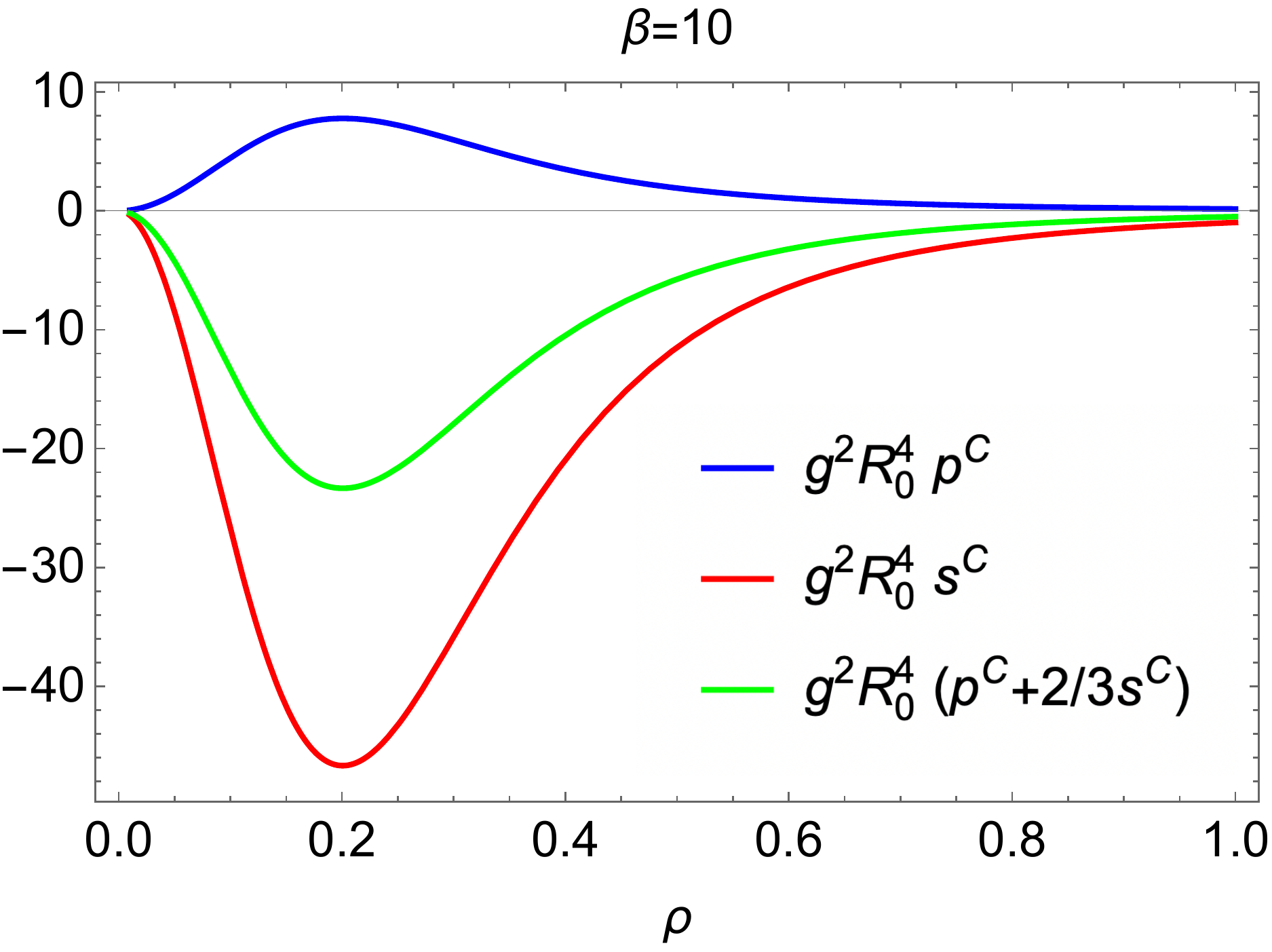}} 
\end{minipage}
\caption{The electromagnetic pressure, share force and normal force distributions as functions of $\rho=r/R_0=gvr$ from Eq.~\eqref{Tcompsnew} for various values of $\beta$.}
\label{vseC}
\end{figure}

We define the electromagnetic EMT analogously to the EMT in electrodynamic as 
\be\label{TmunuC}
T^C_{\mu\nu}=\mathcal{F}_{\mu}^{~\alpha}\mathcal{F}_{\alpha\nu}+\dfrac{1}{4}\eta_{\mu\nu}\mathcal{F}^{\alpha\beta}\mathcal{F}_{\alpha\beta},
\ee
where $\mathcal{F}_{\mu\nu}$ is the $U(1)$ field strength tensor. As we have discussed $\mathcal{F}_{\mu\nu}$ is not uniquely defined. Since we are interesting in spatial structure of the monopole it makes no sense  to define $T^C_{\mu\nu}$  according to the 't Hooft's definition of the $\mathcal{F}_{\mu\nu}$ because of singularities at the origin. Faddeev's and Boulware's smooth definitions of the abelian strength tensor $\mathcal{F}_{\mu\nu}$ could be a better candidate to define the $T^C_{\mu\nu}$, however, we suggest another definition of the $U(1)$ field strength tensor
\be\label{ourF}
\mathcal{F}_{\mu\nu}=\dfrac{\varphi^a}{v}F^a_{\mu\nu}-\dfrac{1}{gv^3}\varepsilon^{abc}\varphi^a\mathcal{D}_{\mu}\varphi^b\mathcal{D}_\nu\varphi^c,
\ee 
which is also smooth as the Faddeev's and Boulware's definitions and it coincides with the 't Hooft's definition at long distances. For this definition of the abelian strength tensor the electromagnetic EMT distributions have the following form
\begin{equation}\label{Tcompsnew}
\begin{split}
T_{00}^{C}(\rho)&=\dfrac{1}{R_0^4g^2}\dfrac{Q^2(\rho)}{2\rho^4},\\
p^{C}(\rho)&=\dfrac{1}{R_0^4g^2}\dfrac{Q^2(\rho)}{6\rho^4},\\
s^{C}(\rho)&=-\dfrac{1}{R_0^4g^2}\dfrac{Q^2(\rho)}{\rho^4},
\end{split}
\end{equation}
where $Q(r)=h(r)\left[1-F^2(r)\left(1-h(r)^2\right)\right]$.  The function $Q(r)$ is directly related to the  magnetic charge density through  
\begin{equation}\label{magchdenful}
\rho_{M}(\rho)=\dfrac{1}{g R_0^3}\dfrac{1}{\rho^2}\dfrac{d}{d\rho}Q(\rho).
\end{equation}   
The electromagnetic EMT densities  are shown in Fig.~\ref{vseC}. The corresponding asymptotic behaviour can be found in App.~\ref{asymptotC}.
Note, that since the full spatial part of the EMT in BSP limit where $\beta=0$ vanishes, see Sec.~\ref{BPSLim}, and the electromagnetic EMT does not depend on $\beta$ (and thereby does not vanish in this limit) the short-range part of the EMT in Eq.~\eqref{TSR} has to be equal to the long-range part. It means that the scalar interaction in BSP limit becomes long-range interacting. 

\subsection{Short-range part of EMT and its consequences}    
\begin{figure}[ht]
\begin{minipage}[h]{0.495\linewidth}
\center{\includegraphics[width=0.87\linewidth]{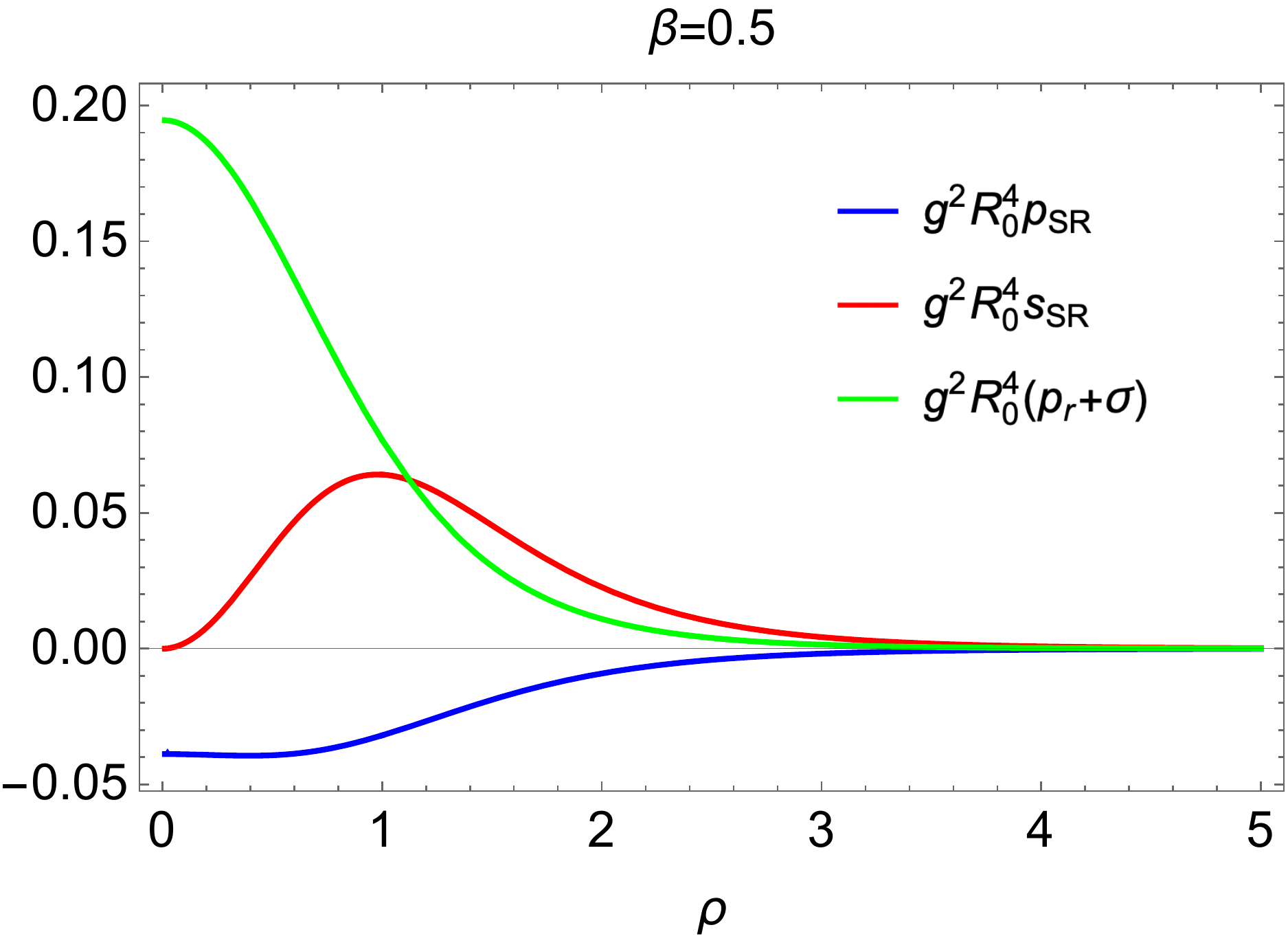}} 
\end{minipage}
\hfill
\begin{minipage}[h]{0.495\linewidth}
\center{\includegraphics[width=0.83\linewidth]{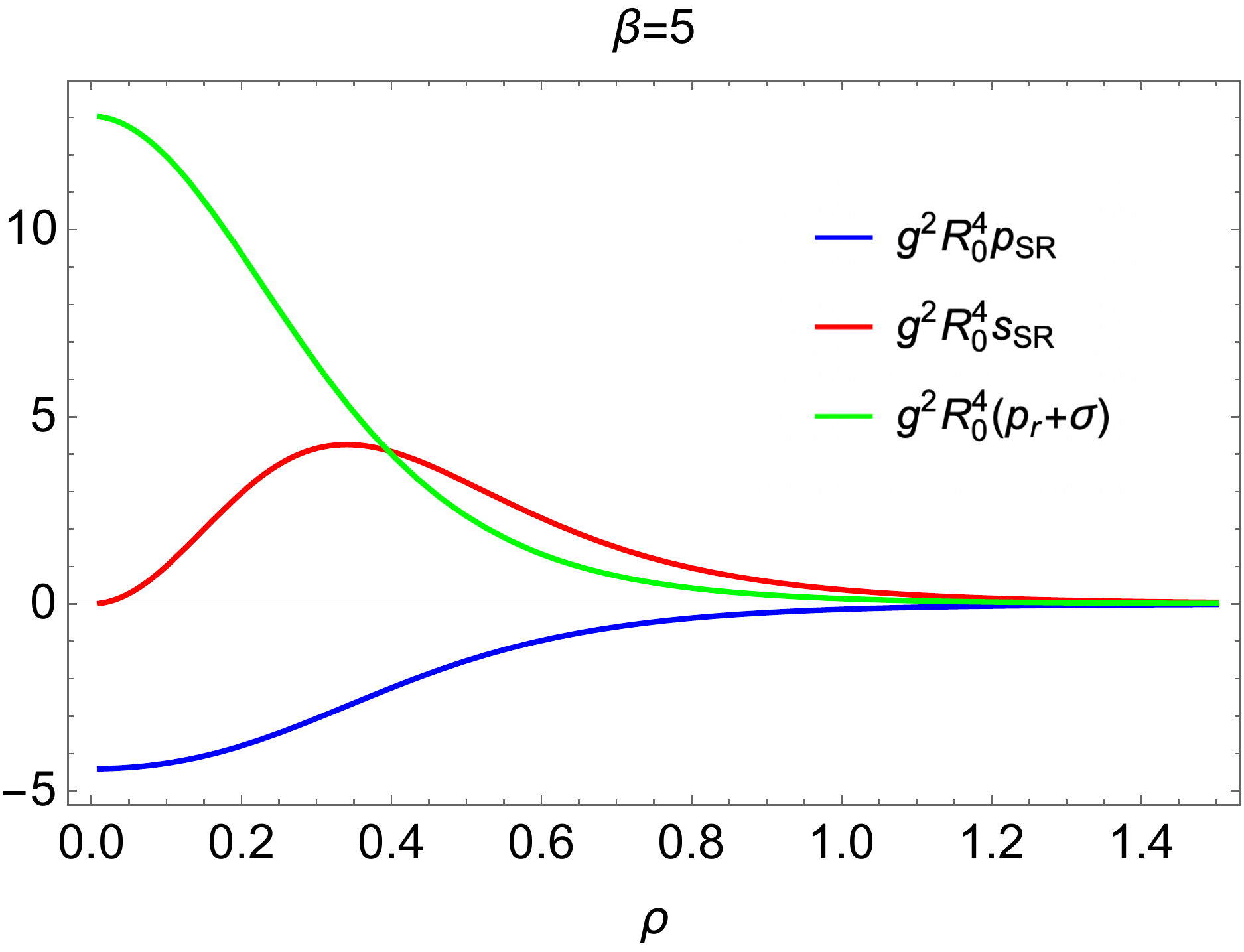}} 
\end{minipage}
\caption{Short-range part of the pressure, shear force and total normal force distributions from Eq.~\eqref{tij_intris}  as a function of $\rho=r/R_0=gvr$ for various values of $\beta$. Note, that the total normal force is defined as in Eq.~\eqref{stabilityJ}.}
\label{short_EMT}
\end{figure}

We denote the short-range part of the EMT as  $T_{\mu\nu}^{SR}$ and define it as follows 
\begin{align}\label{TSR}
T^{SR}_{\mu\nu}=T_{\mu\nu}-T^C_{\mu\nu},
\end{align}
where the full EMT $T_{\mu\nu}$ is obtained   in Eq.~\eqref{energy} and  electromagnetic EMT  -- in Eq.~ \eqref{Tcompsnew}.
After some simple algebraic calculation the following expressions for  the short-range part of the  EMT densities  can be found 
\begin{equation}\label{tij_intris}
\begin{split}
T_{00}^{SR}(\rho)&=\dfrac{1}{R_0^4g^2}\Bigg(\dfrac{F'^2}{\rho^2}+\dfrac{(1-h^2)(1-F^2)^2}{2\rho^4}+\dfrac{1}{2}h'^2+\dfrac{h^2F^2}{\rho^2}\left(1-\dfrac{h^2}{\rho^2}\left[1-F^2\left(1-\dfrac{h^2}{2}\right)\right]\right) + \dfrac{\beta^2}{8}(1-h^2)^2 \Bigg),\\
p^{SR}(\rho)&=\dfrac{1}{R_0^4g^2}\Bigg(\dfrac{1}{3}\dfrac{F'^2}{\rho^2}+\dfrac{(1-h^2)(1-F^2)^2}{6\rho^4}-\dfrac{1}{6}h'^2
-\dfrac{1}{3}\dfrac{F^2h^2}{\rho^2}\left(1+\dfrac{h^2}{\rho^2}\left[1-F^2\left(1-\dfrac{h^2}{2}\right)\right]\right)-\dfrac{\beta^2}{8}(1-h^2)^2\Bigg),\\
s^{SR}(\rho)&=\dfrac{1}{R_0^4g^2}\Bigg(\dfrac{F'^2}{\rho^2}-\dfrac{(1-h^2)(1-F^2)^2}{\rho^4}+h'^2-\dfrac{F^2h^2}{\rho^2}\left(1-2\dfrac{h^2}{\rho^2}\left[1-F^2\left(1-\dfrac{h^2}{2}\right)\right]\right)\Bigg).
\end{split}
\end{equation}
The corresponding asymptotic behaviour can be found in App.~\ref{asymptSR}. The short-range part of the pressure and shear force distributions are presented in Fig.~\ref{short_EMT}. The short-range part of the pressure distribution is  negative and the short-range part of the shear force distribution is  positive for every  choice of $\beta$ in considered range of $r$. The total normal force distribution is positive for any value of $\beta$, which satisfies the stability criterium \eqref{stabilityJ}. It is also interesting to notice that the pressure distribution of the monopole does not change drastically after the subtraction of the long-range part, however, the shear force distribution does.

As we have already discussed in the first section, the static EMT according to the Noether theorem must be conserved.  The conservation of EMT couples pressure and shear force dinsities  by the differential  equation \eqref{StabilityEq}. However, after the  decomposition of the EMT into the short- and long-range parts, the separate EMTs are not conserved and the corresponding short- and long-range parts of the pressure and of the shear force densities couple due to the new equilibrium equations:
\be\label{equilibriumC}
\begin{split}
&\dfrac{d}{dr}\left(\dfrac{2}{3}s^C(r)+p^C(r)\right)+2\dfrac{s^C(r)}{r}=-\dfrac{Q_M(r)\rho_M(r)}{4\pi r^2},\\
&\dfrac{d}{dr}\left(\dfrac{2}{3}s^{SR}(r)+p^{SR}(r)\right)+2\dfrac{s^{SR}(r)}{r}=\dfrac{Q_M(r)\rho_M(r)}{4\pi r^2},
\end{split}
\ee
where $Q_M(r)=\int\limits_{|\vec{x}|<r}d^3x\rho_{M}(\vec{x})=\dfrac{4\pi}{g}Q(r)$ is magnetic charge contained in a sphere of radius $r$. So the right-hand side describes "Coulomb force" of the magnetically charged sphere acting on the magnetic charge density.\footnote{Modulus  of the magnetic field of the magnetically charged sphere is $\mathcal{H}=\dfrac{Q_M(r)}{4\pi r^2}$.}
Thereby the first equation is the equation of magnetostatic equilibrium between the "Coulomb stress" pushing the monopole outwards and the magnetic "Coulomb force" pulling the monopole inward to the center.  In contrast, the second equation describes the balance between the "short-range stress" pulling the monopole inward to the center and  the repulsive magnetic "Coulomb force" pushing the monopole outward. We notice, that since the right-hand sides of the equilibrium equations are associated with the magnetic charge density, it is not possible to define the long- and short-range parts of the EMT using the ambiguity  of the $\mathcal{F}_{\mu\nu}$ in such a way, that the  decomposed  EMTs would be conserved separately, i.e. the right-hand sides would vanish, except the case with zero magnetic charge density.  

As we discussed, the $D$-term of  the monopole diverges because of the long-range  contribution. After  subtracting the long-range part from the EMT, the corresponding $D$-term still diverges for small values of $\beta$.   We illustrate  this divergence in the following. 
The $D$-term is defined through the shear force distribution as in Eq.~\eqref{Dterm1}. Since for small values of $\beta\ll 1$  the main contribution to the short-range part of the shear force density is provided by the asymptotic behaviour at large-distances, see App.~\ref{asymptSR}, the $D$-term diverges for the large enough  $R$  as 
\be\label{Ddiverge}
D\simeq-\dfrac{16\pi M_{SR}R_0}{15g^2}\int\limits_{R}^{\infty}d\rho\rho^4\left(C_h^2\beta^2\dfrac{e^{-2\beta\rho}}{\rho^2}-2C_h\dfrac{e^{-\beta\rho}}{\rho^5}\right)\sim- \int\limits_{R}^{1/\beta}d\rho\left(\beta^2\rho^2-\dfrac{1}{\rho}\right)\sim- \dfrac{1}{\beta}-\ln\beta.
\ee
The mean square magnetic charge radius in Eq.~\eqref{chargeradius}, the mean square energy  radius of the short-range part in Eq.~\eqref{r^2E} and the mean square mechanical  radius in Eq.~\eqref{r^2mech2}  also diverge for small values of $\beta$, see App.~\ref{divR}.   

In the end of this  section we present numerical results for masses, $D$-terms and mean square radii of  the 't Hooft-Polyakov monopole, see Tab.~\ref{nummass-dterm}. We use the following notations: full mass of the monopole $M$ is computed with  help of the energy density from Eq.~\eqref{energy}, electromagnetic contribution to the full mass $M_C$ from Eq.~\eqref{Tcompsnew}, the short-range contribution to the full mass $M_{SH}$ is computed with help of  the expression in Eq.~\eqref{tij_intris}. The values of  $D$-term have been calculated  with help of the short-range part of the shear force distribution in Eq.~\eqref{tij_intris} according to formula \eqref{Dterm1}, the mean square radius of the magnetic charge density is computed due to the definition in Eq.~\eqref{chargeradius},  the mean square radius of the short-range part of the energy distribution is computed according to the definition in Eq.~\eqref{r^2E} and the short-range part of the mechanical square radius -- according to Eq.~\eqref{r^2mech2}. It is remarkable that for the small values of $\beta$ the main contribution to the monopole mass gives the short range part and for the large values of $\beta\gtrsim 5$ -- the long range part. It would be interesting to find the analytic form of the pressure and shear force densities depending on  the parameter $\beta$ like it was done in Refs. \cite{Forgacs:2005vx,Gardner:1982fk,Kirkman:1981ck}  for the static energy. 
 \begin{table}[h]
                           \begin{center}
                           \begin{tabular}{c|c|c|c|c|c|c|c}
                           $\beta$ & $M \dfrac{R_0g^2}{4\pi}$ & $M_C\dfrac{R_0g^2}{4\pi}$ &$M_{SR}\dfrac{R_0g^2}{4\pi}$ &  $-\dfrac{15g^4}{64\pi^2}D$ &  $\dfrac{\langle r^2\rangle_{M}}{R_0^2}$ &   $\dfrac{\langle r^2\rangle_{E}}{R_0^2}$  & $\dfrac{\langle r^2\rangle_{\text{mech}}}{R_0^2}$\\ \hline
                          $\varepsilon$& 0.993& 0.153& 0.841&  $ \infty$ &$\infty$ &$\infty$&$\infty$   \\
                        0.1& 1.036& 0.187& 0.849& 7.909 & 20.517&  12.157&  20.255 \\  
                          0.5& 1.133& 0.267& 0.865& 1.022 &  4.287&   3.710& 3.379 \\
                         1&1.204& 0.336& 0.869& 0.43&  2.168&  2.436& 1.545 \\
                          5&  1.438& 0.794& 0.644& 0.293 &  0.373&    1.602& 0.445 \\
                      8& 1.457& 1.084& 0.373 & 0.213 &0.205 &2.328 &0.307 \\
                          10 &1.461& 1.289& 0.172&0.108&  0.151&    4.479&  0.254  \\
           
                           \end{tabular}
                           \caption{Numerical values of $D$-terms, various radii and different contributions to the mass for varied choice of $ \beta$, where $\varepsilon\ll 1$. The estimation of the accuracy of our numerical method is  about 95 percent  for $0\leq\beta\leq1$ and about 85 percent for $\beta> 1$. }
                           \label{nummass-dterm}
                           \end{center}
                           \end{table}                        

\newpage
 \section{Julia-Zee dyon}
 \label{dyon}
 \subsection{Equation of motion}
In the previous section we considered monopole solution carrying  magnetic charge. In this part we will consider a dyon, that  is a hypothetical particle carrying  both electric and magnetic charges, such particle was first suggested  by Julian Schwinger  in 1969 in Ref.~\cite{Schwinger:1969ib}. The dyon solution in Georgi-Glashow model was first obtained in 1975 by Anthony Zee and   Bernard Julia in Ref.~\cite{Julia:1975ff}.
We will use the same Georgi-Glashow model as we did in the first part with the same action \eqref{action}. However, we will not fix the zero component of the vector field as we did it for the monopole case, $A^a_0(\vec{x})\neq 0$. We are again interested in the static soliton solution with finite energy. Thereby, we choose  the same spherically-symmetric ansatz as we chose for the 't Hooft-Polyakov monopole in Eq.~\eqref{anzatz} with the same boundary condition as in Eq.~\eqref{boundcon}. In contrast to monopole case, for the dyon the  $A_0^a$ is non-zero and one chooses the  ansatz 
\begin{align}\label{ansatzDyon}
A^a_0&=\dfrac{J(r)}{gr}\dfrac{r^a}{r},
\end{align}
which follows from requiring the energy to be  finite and definition of the abelian field strength tensor as in Eq.~\eqref{ourF}. The boundary conditions for the function $J(r)$ are
\be
\begin{split}
J(r)&\underset{r\to 0}\simeq 0,\\
J(r)&\underset{r\to \infty}\simeq -\dfrac{Q_D}{Q_M} +m g r,
\end{split} 
\ee
where $Q_D$ is an electric charge of a dyon, $Q_M=\dfrac{4\pi}{g}$ is a monopole charge and $m$ is a free constant, that has dimension of mass.   

From the variation of the action follow the equations of motion 
\begin{equation}
\begin{split}
&\mathcal{D}_0F^a_{00}-\mathcal{D}_iF^a_{i0}+g\epsilon^{abc}\varphi^b\mathcal{D}_0\varphi^c=0,\\
&\mathcal{D}_0F^a_{0k}-\mathcal{D}_iF^a_{ik}-g\epsilon^{abc}\varphi^c\mathcal{D}_k\varphi^b=0,\\
&\mathcal{D}_0\mathcal{D}_0\varphi^a-\mathcal{D}_i\mathcal{D}_i\varphi^a+\lambda(\varphi^b\varphi^b-v^2)\varphi^a=0.
\end{split}
\end{equation} 
The equations for the profile functions in  dimensionless variable $\rho=g v r$ are \begin{equation}\label{dyoneqs}
\begin{split}
&F''\rho^2=\left(F^2-1\right)F-\left(J^2-\rho^2h^2\right)F,\\
&h''\rho^2+2h'\rho=2F^2h+\dfrac{\beta^2}{2}h\left(h^2-1\right)\rho^2,\\
&J''\rho^2=2JF^2.
\end{split}
\end{equation}
Note, that the boundary condition for $J(\rho)$ also changes:  $J(\rho)\underset{\rho\to \infty}\simeq -\dfrac{Q_D}{Q_M} +C \rho $, where $C=m/v $ is dimensionless free constant.

The approximate solution for small and large distances can be  found:
\begin{equation}\label{eqmorigin}
\begin{split}
F(\rho)&\underset{\rho\to 0}\simeq  1+a\rho^2+\sum_{n=2}^{\infty}a_{2n}\rho^{2n},\hspace{8pt} F(\rho)\underset{\rho\to \infty}\simeq C_Fe^{-\sqrt{1-C^2}\rho}\left(1+O\left(\dfrac{1}{\rho}\right)\right),\\
h(\rho)&\underset{\rho\to 0}\simeq  b \rho+\sum_{n=1}^{\infty}b_{2n+1}\rho^{2n+1},\hspace{8pt} h(\rho)\underset{\rho\to \infty}\simeq 1-C_h\dfrac{e^{-\beta\rho}}{\rho}\left(1+O\left(\dfrac{1}{\rho}\right)\right)-\dfrac{2C_F^2}{\beta^2+4C^2-4}\dfrac{e^{-2\sqrt{1-C^2}\rho}}{\rho^2}\left(1+O\left(\dfrac{1}{\rho}\right)\right),\\
\tilde{J}(\rho)&\underset{\rho\to 0}\simeq  c\rho+\sum_{n=1}c_{2n+1}\rho^{2n+1},\hspace{8pt}\tilde{J}(\rho)\underset{\rho\to \infty}\simeq  C-\dfrac{Q_D}{Q_M}\dfrac{1}{\rho}+\dfrac{C C_F^2}{2(1-C^2)}\dfrac{e^{-2\sqrt{1-C^2}\rho}}{\rho^2}\left(1+O\left(\dfrac{1}{\rho}\right)\right).
\end{split}
\end{equation}
The  $a,\ b,\ c$ are free constants, all other constants for the small $r$ behaviour  can be expressed in terms of these constants, for example,
\begin{equation}
a_4=\dfrac{1}{10}\left(3a^2+b^2-c^2\right),\hspace{10pt}b_3=\dfrac{1}{10}\left(4 ab-\dfrac{b\beta^2}{2}\right),\hspace{10pt}c_3=\dfrac{2ac}{5}.
\end{equation}
Since we searched the asymptotic behaviour at large distances  for real and falling functions we define a new function $J(\rho)=\tilde{J}(\rho)\rho$ and find that  the constant $C$ has restricted region: $0 \leq C<1$. From numerical analysis of the solutions we obtain that the charge ratio is restricted in the following range $0\leq \dfrac{Q_D}{Q_M}\leq 1$. More detailed analysis can be found in Ref.~\cite{Nishino:2019lqm}, where also  the dependence of the parameter $C$ on the charge ratio $\dfrac{Q_D}{Q_M}$ is found. 

\subsection{EMT densities}
\begin{figure}[h]
\begin{minipage}[h]{0.3\linewidth}
\center{\includegraphics[width=1\linewidth]{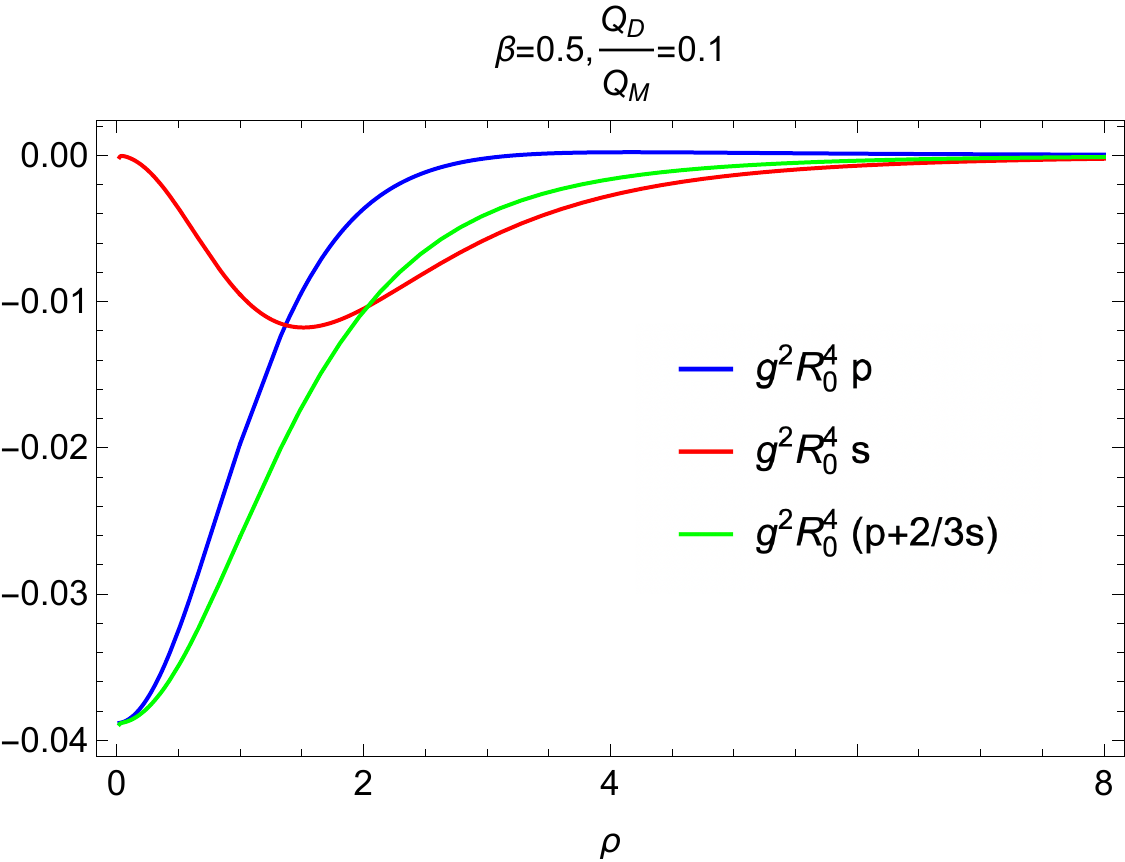}} \\
\end{minipage}
\hfill
\begin{minipage}[h]{0.3\linewidth}
\center{\includegraphics[width=1\linewidth]{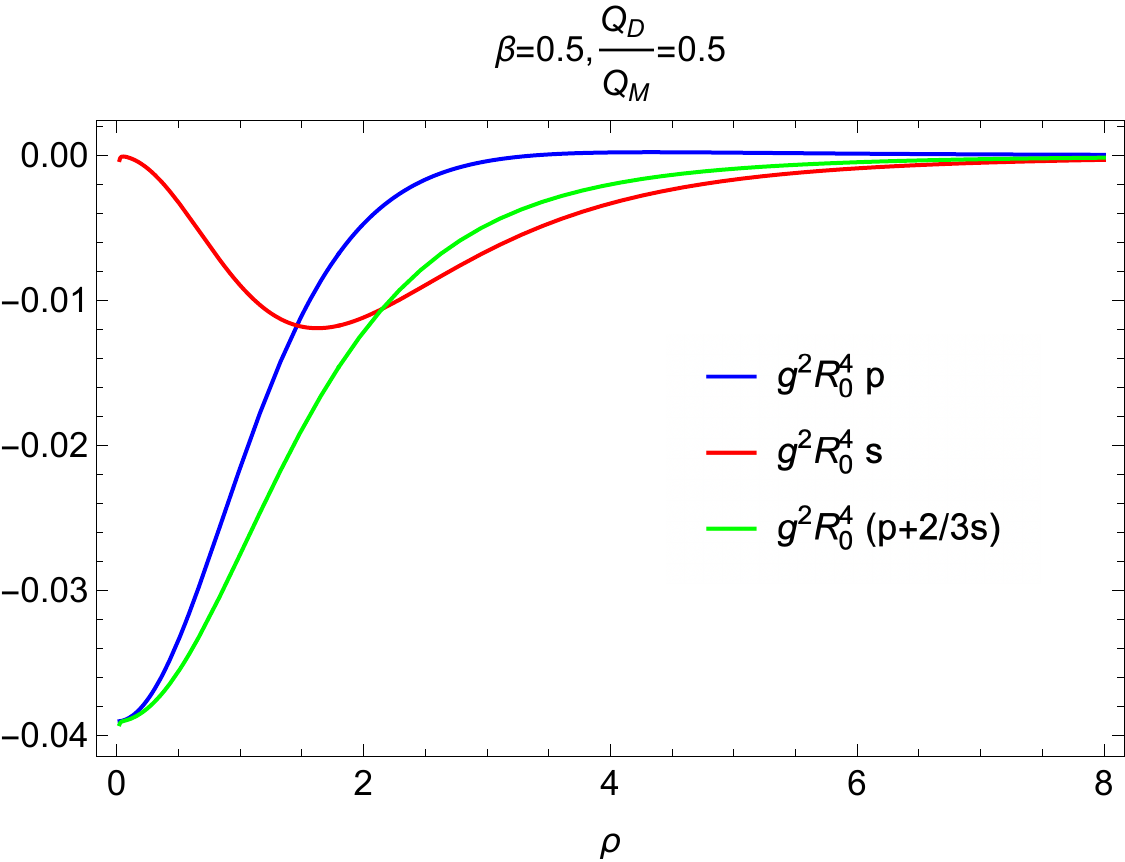}}\\
\end{minipage}
\hfill
\begin{minipage}[h]{0.3\linewidth}
\center{\includegraphics[width=1\linewidth]{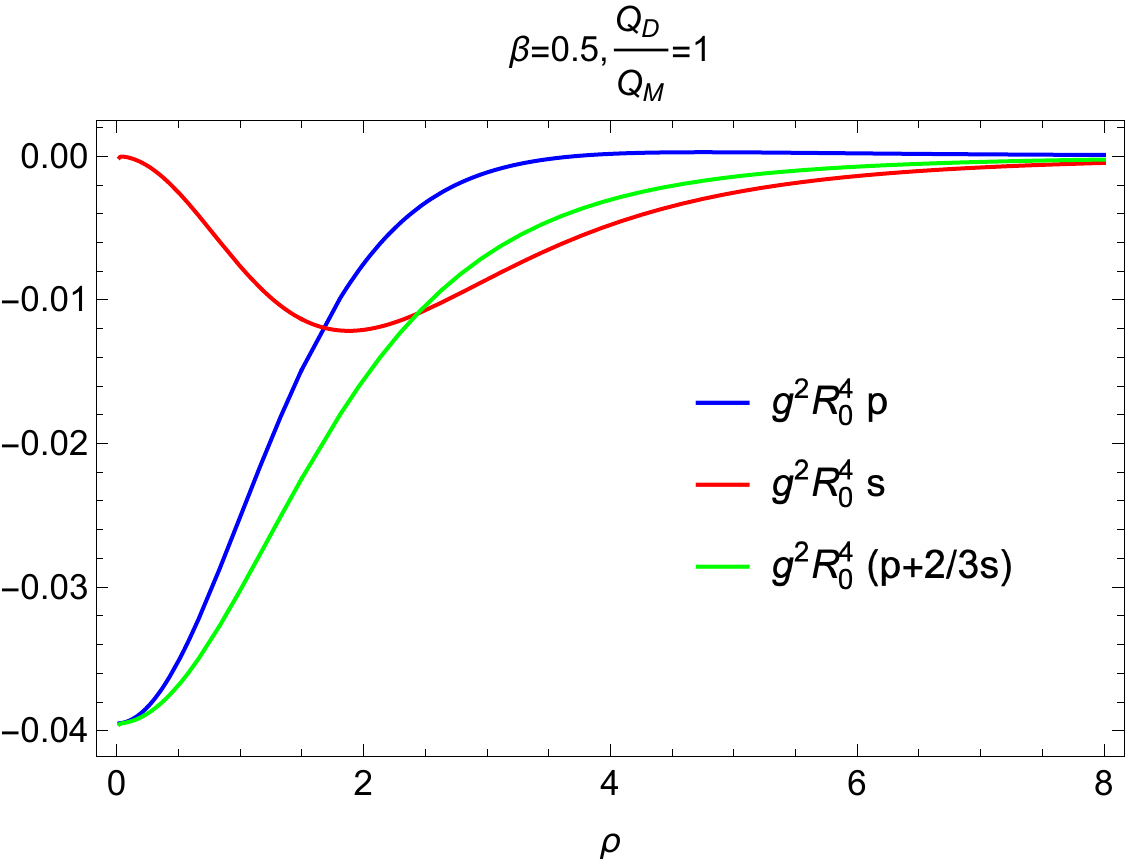}}\\
\end{minipage}
\vfill
\begin{minipage}[h]{0.3\linewidth}
\center{\includegraphics[width=1\linewidth]{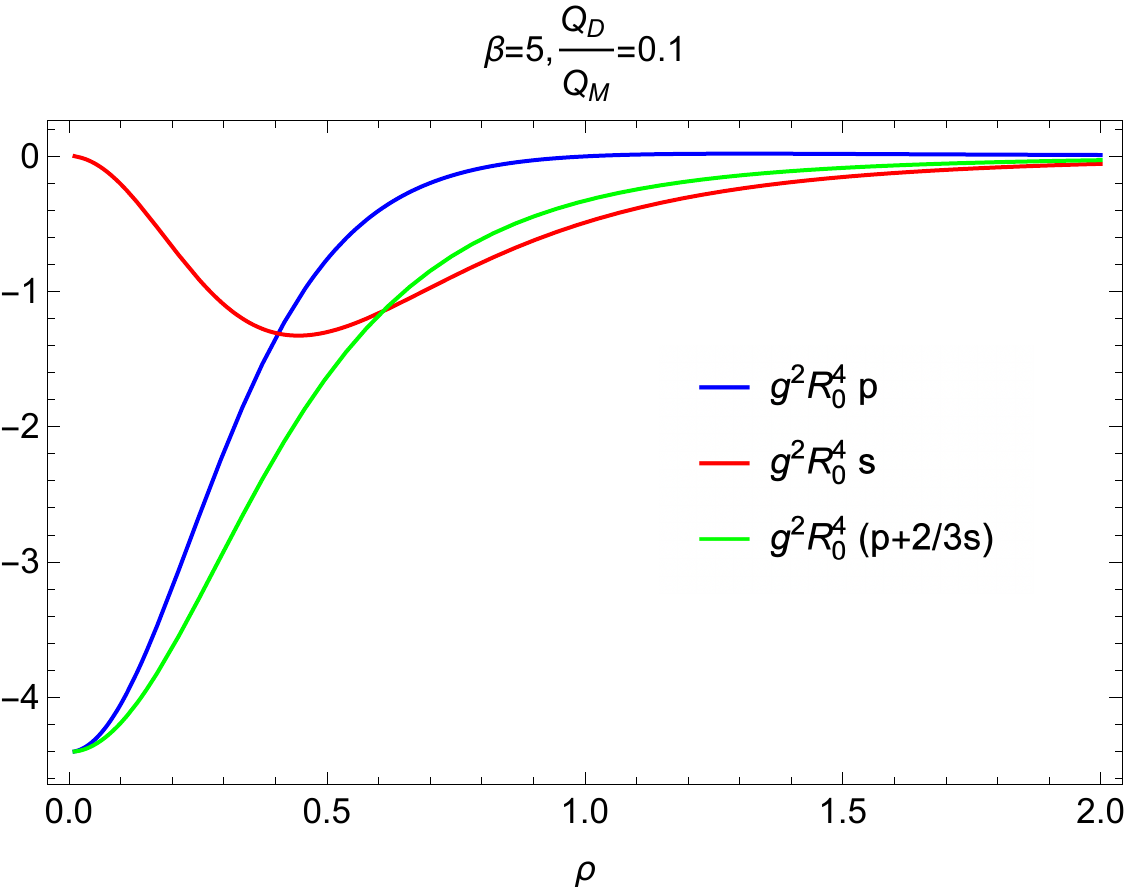}}  \\
\end{minipage}
\hfill
\begin{minipage}[h]{0.3\linewidth}
\center{\includegraphics[width=1\linewidth]{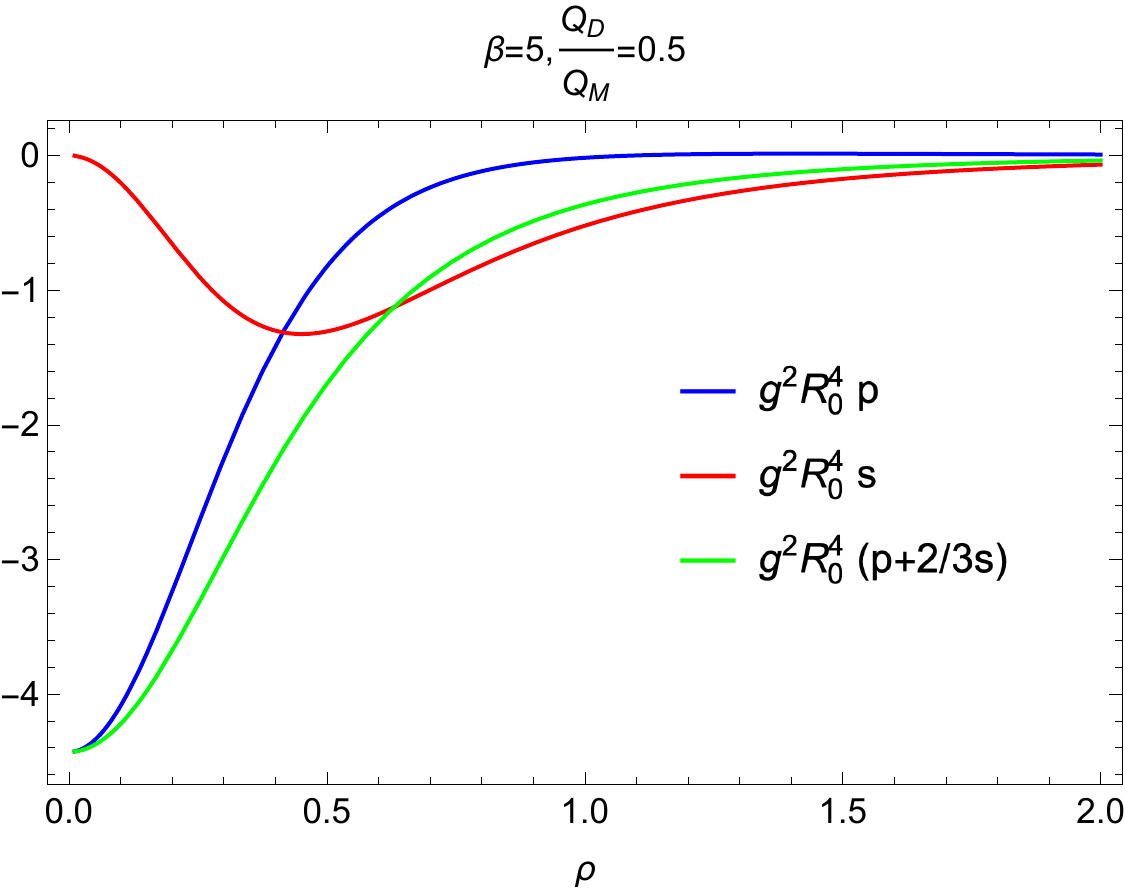}} \\
\end{minipage}
\hfill
\begin{minipage}[h]{0.3\linewidth}
\center{\includegraphics[width=1\linewidth]{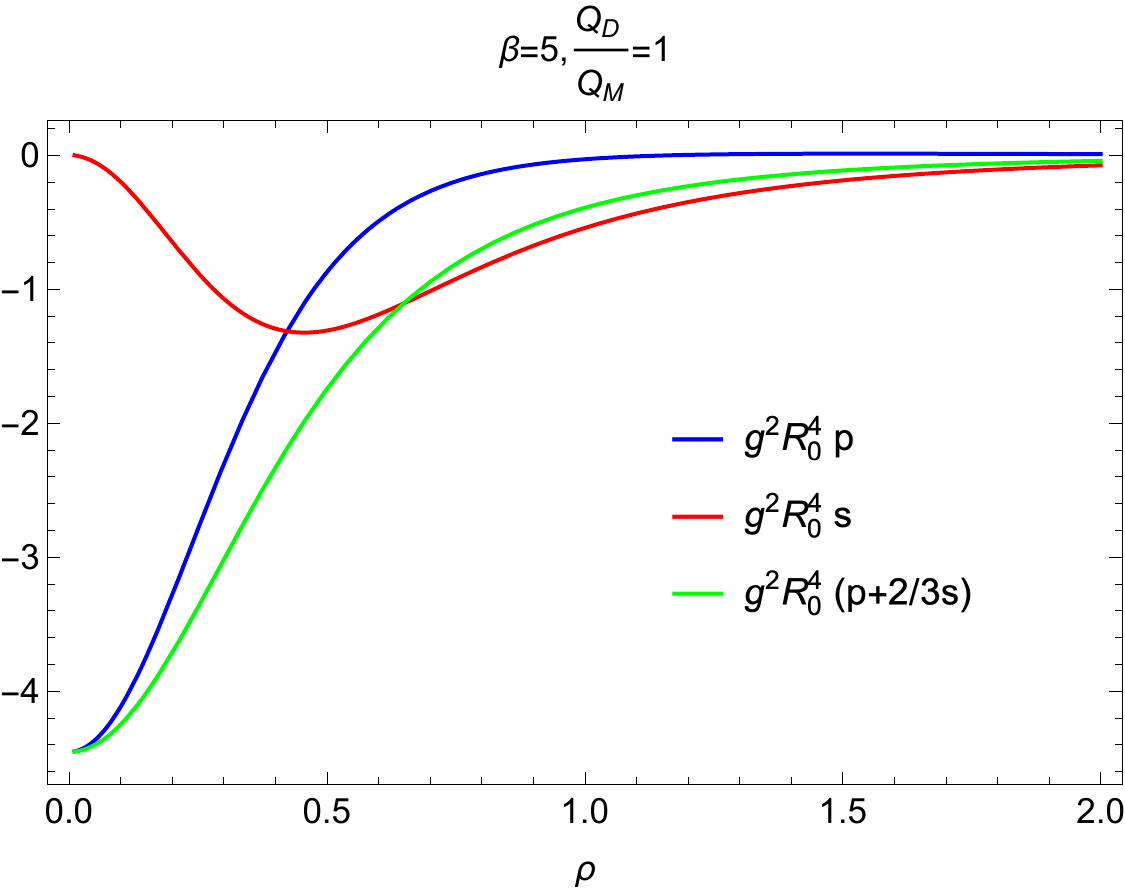}} \\
\end{minipage}
\caption{The mechanical properties of dyons from Eq.~\eqref{T00d} as functions of $\rho=r/R_0=gvr$ for varied value of $\beta$ and charge ratio $\dfrac{Q_D}{Q_M}$. }
\label{DyonDensityold}
\end{figure}

\begin{figure}[ht]
\begin{minipage}[h]{0.495\linewidth}
\center{\includegraphics[width=0.95\linewidth]{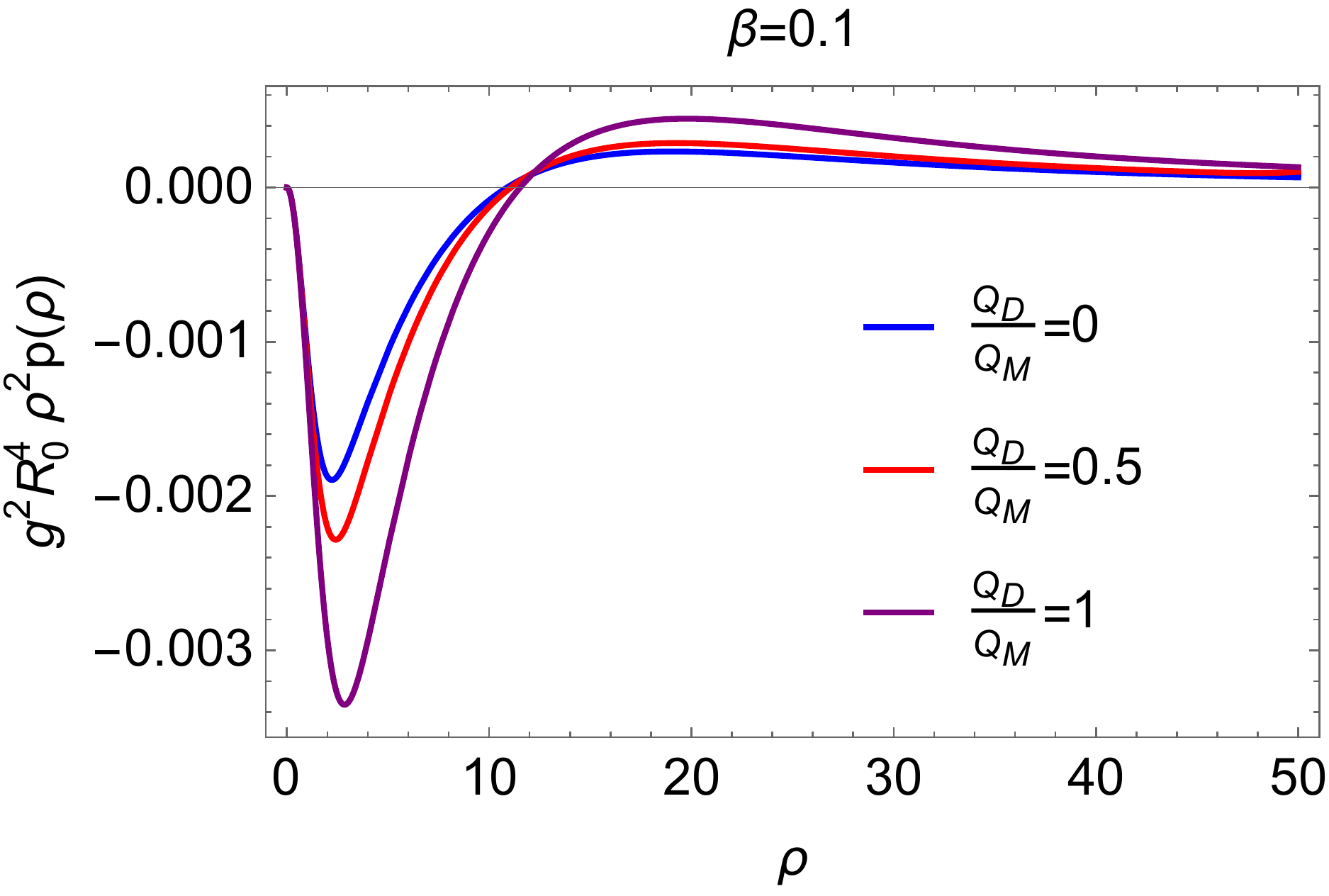}} \\
\end{minipage}
\hfill
\begin{minipage}[h]{0.495\linewidth}
\center{\includegraphics[width=0.95\linewidth]{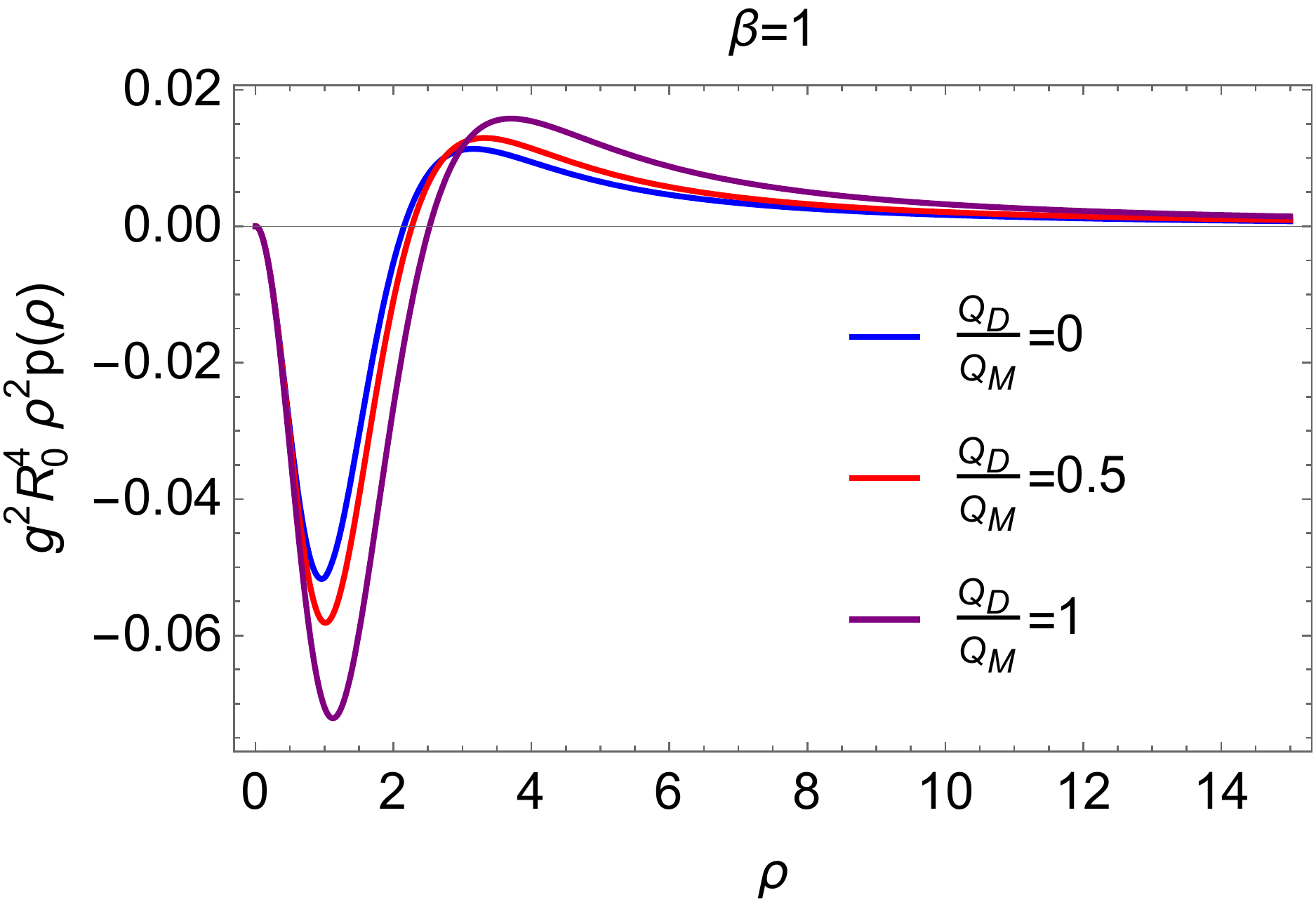}} \\
\end{minipage}
\vfill
\caption{The pressure distributions of the monopole $\left(\dfrac{Q_D}{Q_M}=0\right)$ and the dyon  for various values of $\beta$ and charge ratio $\dfrac{Q_D}{Q_M}$.  The pressure distribution has one mode for any choice of the parameters.}
\label{PressureNul}
\end{figure}

\begin{figure}[ht]
\begin{minipage}[h]{0.495\linewidth}
\center{\includegraphics[width=0.75\linewidth]{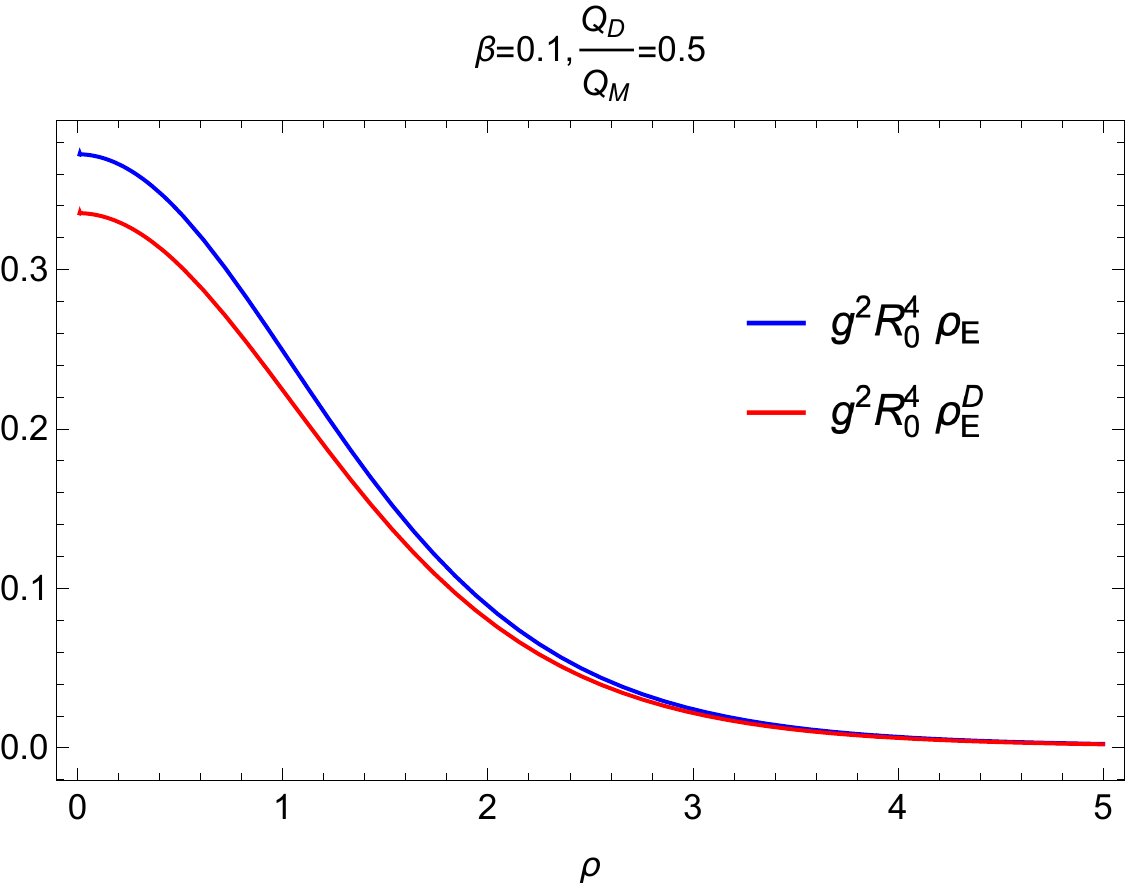}} \\
\end{minipage}
\hfill
\begin{minipage}[h]{0.495\linewidth}
\center{\includegraphics[width=0.75\linewidth]{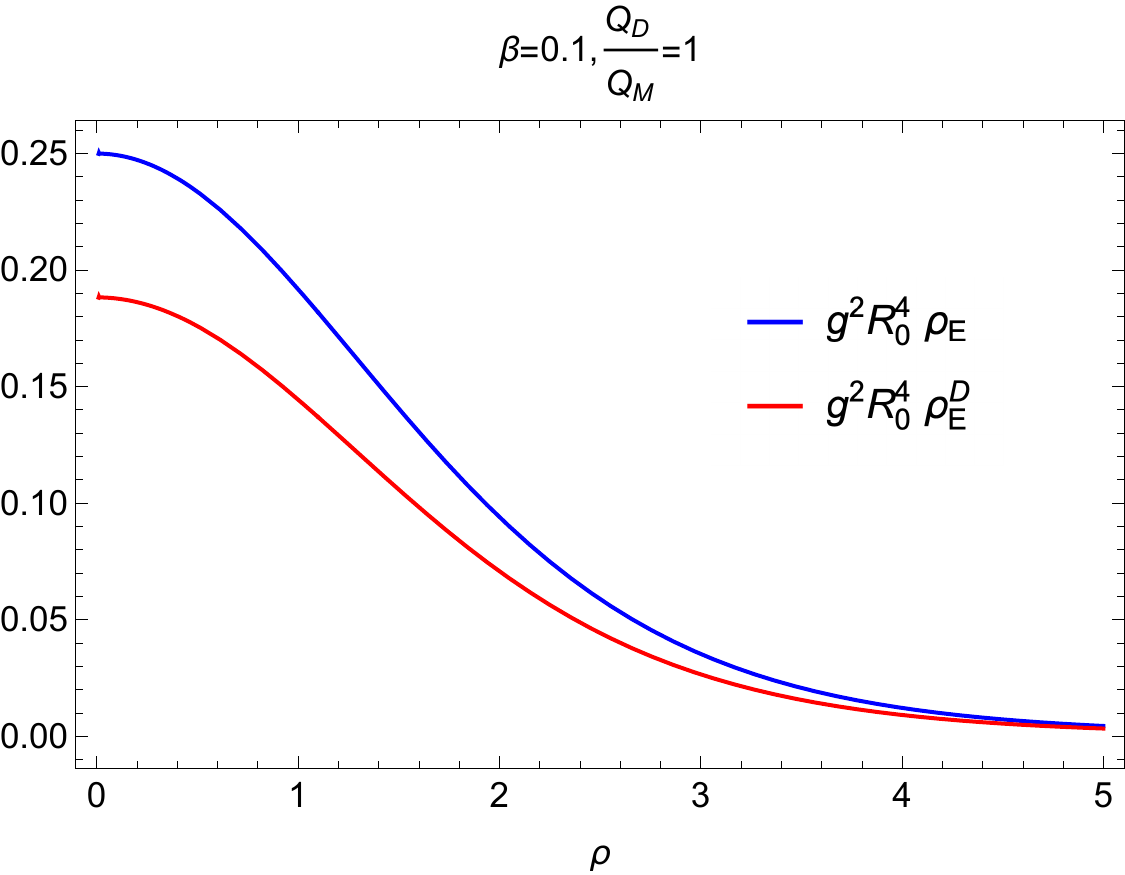}} \\
\end{minipage}
\vfill
\begin{minipage}[h]{0.495\linewidth}
\center{\includegraphics[width=0.75\linewidth]{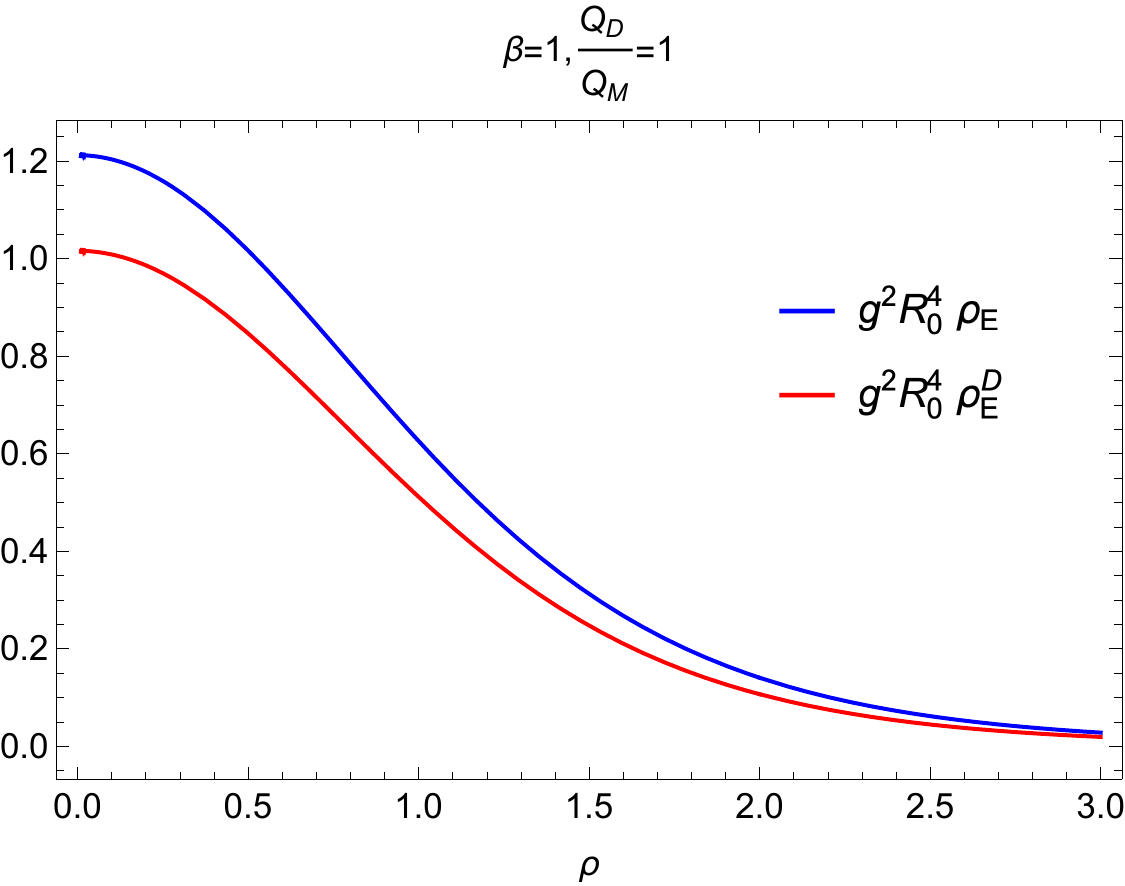}} \\
\end{minipage}
\hfill
\begin{minipage}[h]{0.495\linewidth}
\center{\includegraphics[width=0.75\linewidth]{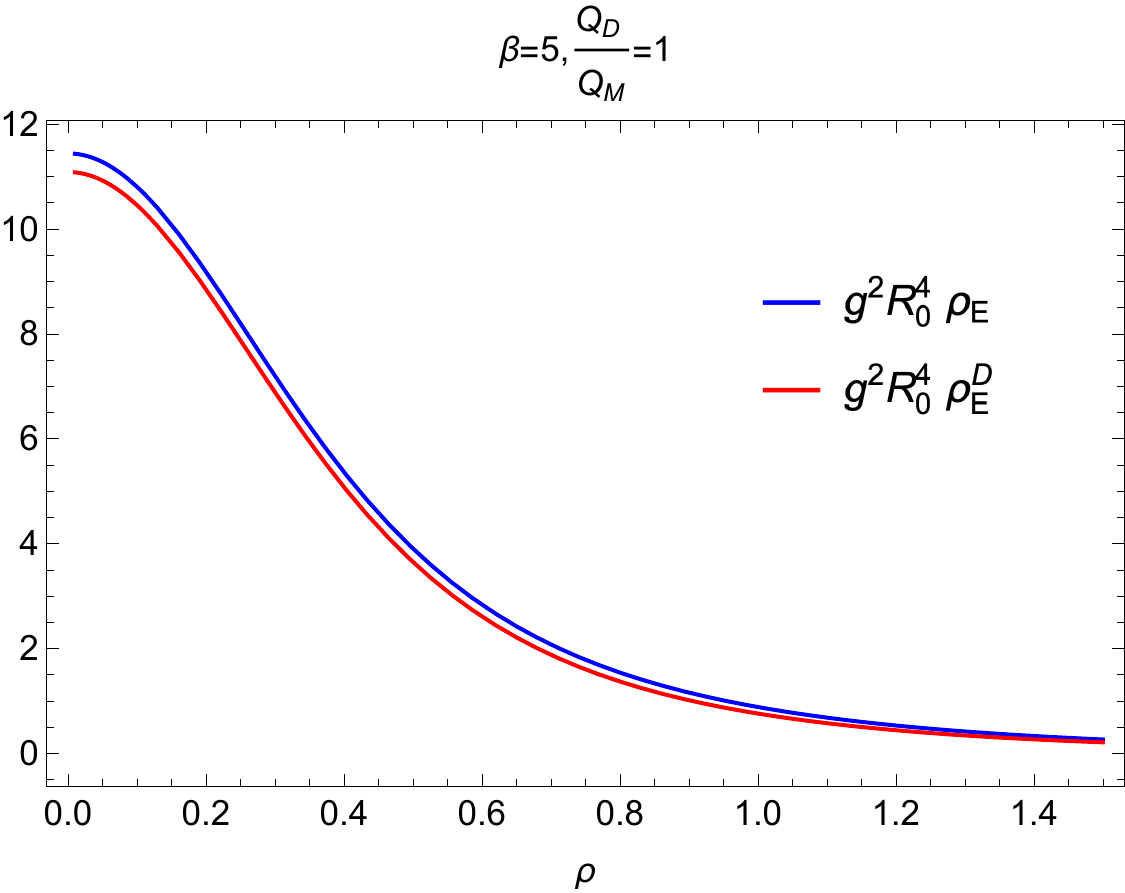}} \\
\end{minipage}
\caption{Comparison of the full energy densities of dyon and the contributions of the monopole part only  for various values of $\beta$ and charge ratio $\dfrac{Q_D}{Q_M}$.}
\label{DyonEnergy}
\end{figure}

The EMT for the Julia-Zee dyon can be obtained analogically to the 't Hooft-Polyakov monopole by variating generally covariant form of the Georgi-Glashow action \eqref{action} with respect to the metric and it  can be decomposed into magnetic and electric parts:
\begin{equation}\label{T00d}
\begin{split}
T_{00}(r)&=T_{00}^M(r)+T_{00}^{E}(r),\\
T_{ij}(r)&=T_{ij}^M(r)+T_{ij}^E(r),
\end{split}
\end{equation}
where the magnetic part equals to the EMT of monopole  which we have already computed in the previous section, see Eqs.~\eqref{T00}, \eqref{energy} and \eqref{EMTinf}. The electric part of EMT is 
\begin{equation}
\begin{split}
T_{00}^E(r)&=\dfrac{1}{2}\left(\mathcal{D}_0\varphi^a\mathcal{D}_0\varphi^a+F^a_{0i}F^a_{0i}\right),\\
T_{ij}^E(r)&=\dfrac{1}{2}\left[\delta_{ij}\left(F^a_{0k}F^a_{0k}+\mathcal{D}_0\varphi^a\mathcal{D}_0\varphi^a\right)-2F^a_{i0}F^a_{j0}\right].
\end{split}
\end{equation}
Note, that same as for the monopole case the spatial components of the EMT for the dyon vanish in BSP limit, where $\beta=0$.  
For electric part of the energy density, pressure and shear force distributions of the dyon   in dimensionless variable $\rho$ the following expressions are obtained
\begin{equation}\label{Tijdyon2}
\begin{split}
&T_{00}^E(\rho)=\dfrac{1}{R_0^4g^2}\left(\dfrac{\tilde{J}'^2}{2}+\dfrac{\tilde{J}^2F^2}{\rho^2}\right),\\
p^E(\rho)=\dfrac{1}{R_0^4g^2}&\left(\dfrac{\tilde{J}'^2}{6}+\dfrac{\tilde{J}^2F^2}{3\rho^2}\right),\hspace{15pt}s^E(\rho)=\dfrac{1}{R_0^4g^2}\left(-\tilde{J}'^2+\dfrac{\tilde{J}^2F^2}{\rho^2}\right).
\end{split}
\end{equation}
With the help of asymptotic behaviour of  profile functions in Eq.~\eqref{eqmorigin} the asymptotic behaviour of the EMT for the dyon  can be obtained
\be  
\begin{split}\label{emtDinf}
&T_{00}(\rho)\underset{\rho\to \infty}\simeq\dfrac{1}{R_0^4g^2}\Bigg[ \dfrac{1}{2}\left(1+\left(\dfrac{Q_D}{Q_M}\right)^2\right)\dfrac{1}{\rho^4}+2C_F^2\dfrac{e^{-2\sqrt{1-C^2}\rho}}{\rho^2}\left(1+O\left(\dfrac{1}{\rho}\right)\right)+\beta^2C_F^2\dfrac{e^{-2\beta\rho}}{\rho^2}\left(1+O\left(\dfrac{1}{\rho}\right)\right)\Bigg], \\
&p(\rho)\underset{\rho\to \infty}\simeq\dfrac{1}{R_0^4g^2}\Bigg[\dfrac{1}{6}\left(1+ \left(\dfrac{Q_D}{Q_M}\right)^2\right)\dfrac{1}{\rho^4}-\dfrac{2}{3}\beta^2C_F^2\dfrac{e^{-2\beta\rho}}{\rho^2}\left(1+O\left(\dfrac{1}{\rho}\right)\right)+O\left(\dfrac{e^{-2\sqrt{1-C^2}\rho}}{\rho^3}\right)\Bigg],\\
&s(\rho)\underset{\rho\to \infty}\simeq \dfrac{1}{R_0^4g^2}\Bigg[-\left(1+\left(\dfrac{Q_D}{Q_M}\right)^2\right)\dfrac{1}{\rho^4}+\beta^2C_F^2\dfrac{e^{-2\beta\rho}}{\rho^2}\left(1+O\left(\dfrac{1}{\rho}\right)\right)+O\left(\dfrac{e^{-2\sqrt{1-C^2}\rho}}{\rho^3}\right)\Bigg].
\end{split}
\ee 
From this behaviour one sees that the long-range contribution presented  in the dyon  is even stronger than in the monopole case, see Eq.~\eqref{EMTinf}.
In App.~\ref{DyonApp} one can find the asymptotic behaviour of the EMT near the origin as well as behaviour of the electric part of the EMT only. Again already from the asymptotic behaviour it is clear that  the stability condition of Eq.~\eqref{Mstability} is violated.
In Fig.~\ref{DyonDensityold} the full pressure and shear force distributions of the dyon are presented. The shear force and pressure distributions have very similar form as it was in the monopole case, see Fig.~\ref{oldemt}. These distributions  are mostly negative for every choice of parameters. However, pressure distribution changes its sing, what allows the von Laue condition \eqref{Laucond} to be satisfied, see e.g. Fig.~\ref{PressureNul}.  It is also interesting to notice that the main contribution to the dyon energy comes from the monopole part as it can be seen in  Fig.~\ref{DyonEnergy}. Moreover, for the growing  $\beta$ this contribution is increasing. 

\subsection{Electromagnetic and short-range parts of the EMT}
According to the definition of the electromagnetic EMT in Eq.~\eqref{TmunuC} the long-range part of the EMT for dyon is
  \begin{equation}
  \begin{split}
T_{00}^{C}&=\dfrac{1}{4}\mathcal{F}_{ij}\mathcal{F}_{ij}+\dfrac{1}{2}\mathcal{F}_{0i}\mathcal{F}_{0i},\\
T_{ij}^{C}&=-\dfrac{1}{4}\delta_{ij}\mathcal{F}_{mn}\mathcal{F}_{mn}-\mathcal{F}_{ik}\mathcal{F}_{kj}+\dfrac{1}{2}\delta_{ij}\mathcal{F}_{0k}\mathcal{F}_{0k}-\mathcal{F}_{i0}\mathcal{F}_{j0}.
  \end{split}
  \end{equation}
 Using  the modified 't Hooft abelian field strength tensor in Eq.~\eqref{ourF} one gets the following expressions for the electromagnetic EMT densities of a dyon
 \begin{equation}\label{coulombful}
 \begin{split}
 T_{00}^{C}(\rho)&=\dfrac{1}{2}\dfrac{1}{g^2R_0^4}\dfrac{1}{\rho^4}\left(Q^2(\rho)+\tilde{Q}^2(\rho)\right),\\
 p^{C}(\rho)&=\dfrac{1}{6}\dfrac{1}{g^2R_0^4}\dfrac{1}{\rho^4}\left(Q^2(\rho)+\tilde{Q}^2(\rho)\right),\\
 s^{C}(\rho)&=-\dfrac{1}{g^2R_0^4}\dfrac{1}{\rho^4}\left(Q^2(\rho)+\tilde{Q}^2(\rho)\right),
 \end{split}
 \end{equation} 
 where the function $Q$ is defined in Eq.~\eqref{Tcompsnew} and is related to the magnetic charge density in Eq.~\eqref{magchdenful} and  the function $\tilde{Q}$ is related to the electric charge density of a dyon 
 \begin{equation}\label{eldistr}
\rho_D(\rho)=\dfrac{1}{gR_0^3}\dfrac{1}{\rho^2}\dfrac{d\tilde{Q}(\rho)}{d\rho}, \hspace{12pt}\text{with}\hspace{12pt}\tilde{Q}(\rho)=\rho^2h(\rho)\tilde{J}'(\rho).
\end{equation}
Comparing  these expressions with the expressions for the  monopole case in Eq.~\eqref{Tcompsnew}, it is clear that both have similar behaviour. Hence, it is again not possible  to define the various mean square radii or $D$-terms, because they diverge. 

\begin{figure}[ht]
\begin{minipage}[h]{0.495\linewidth}
\center{\includegraphics[width=0.85\linewidth]{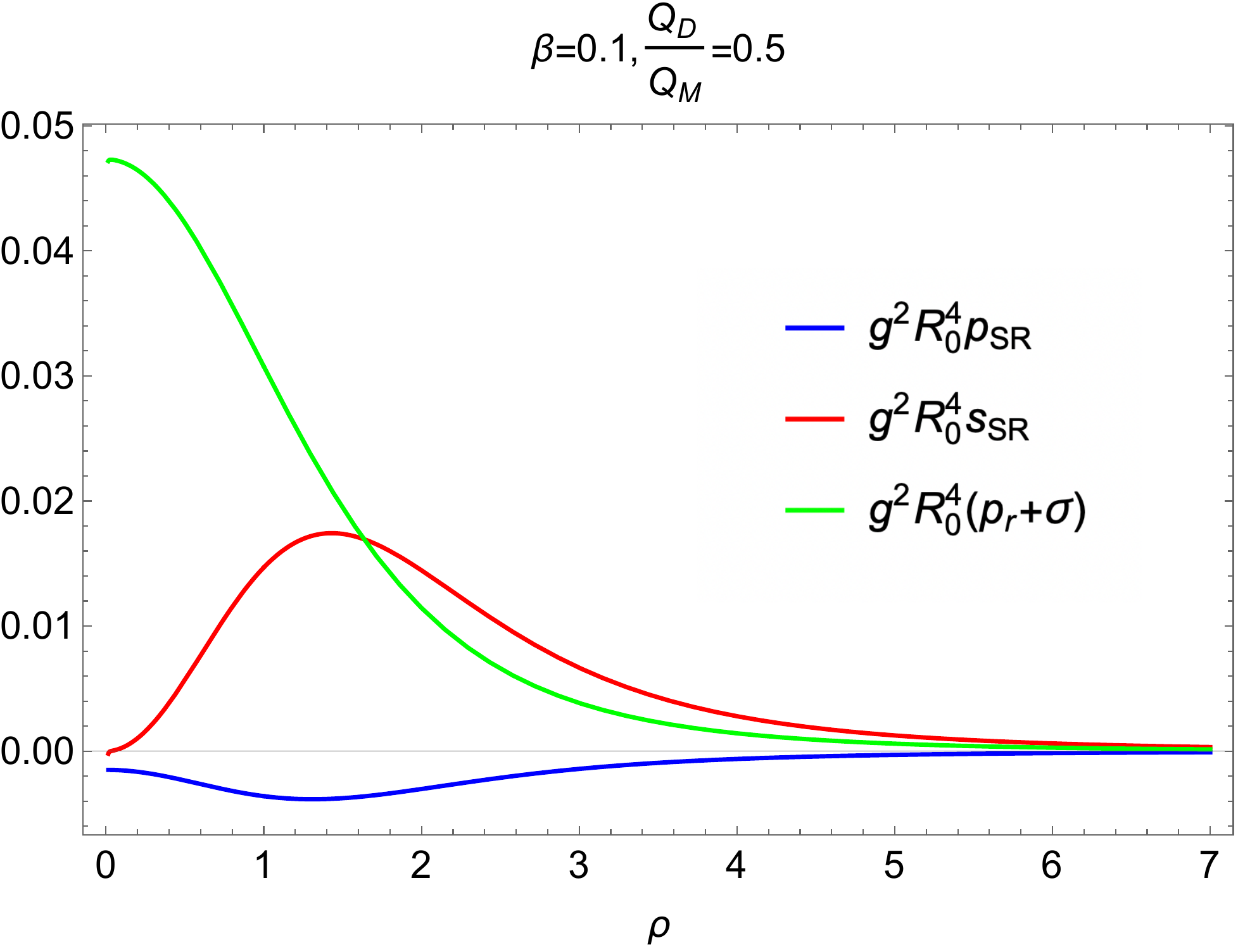}} \\
\end{minipage}
\hfill
\begin{minipage}[h]{0.495\linewidth}
\center{\includegraphics[width=0.85\linewidth]{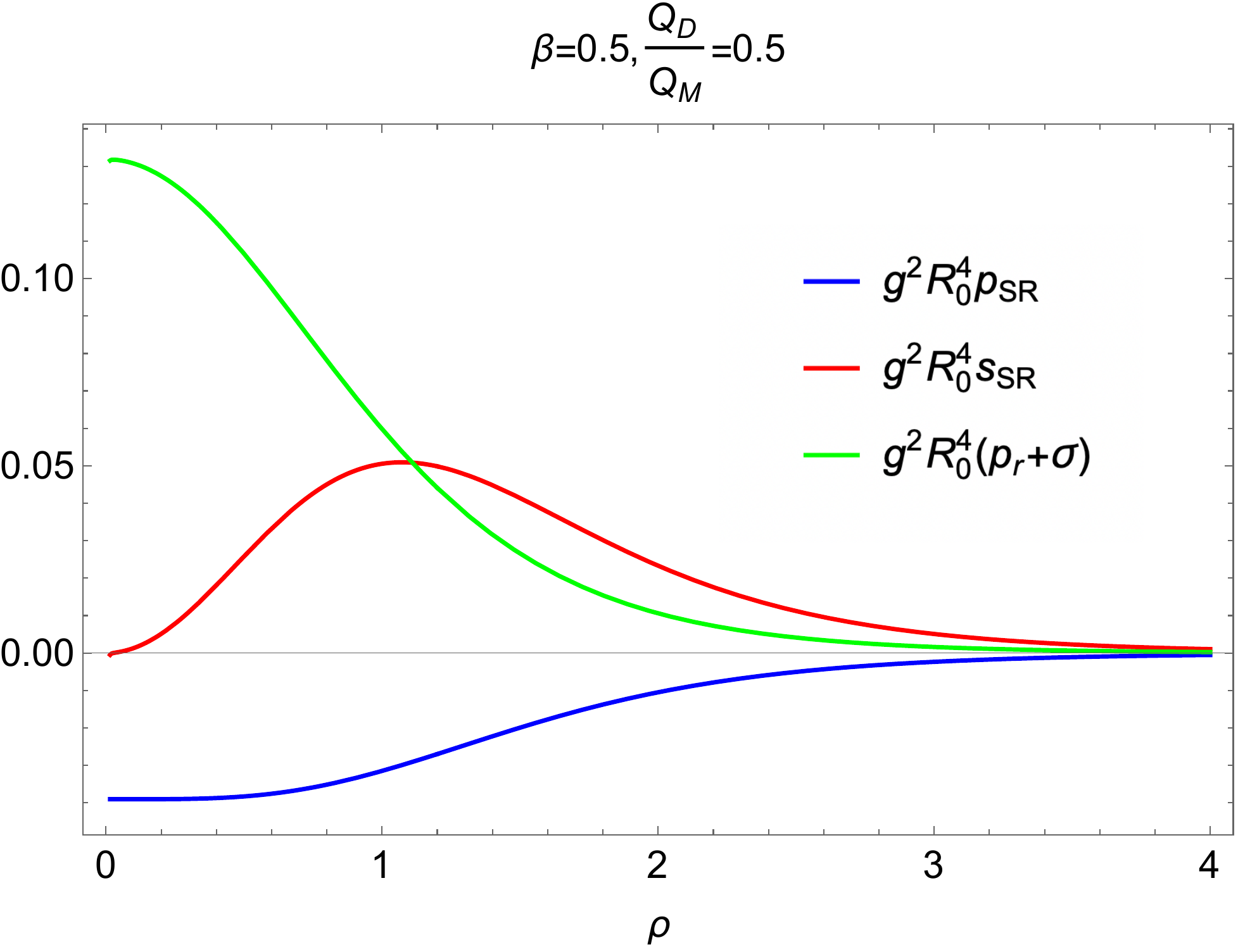}} \\
\end{minipage}
\vfill
\begin{minipage}[h]{0.495\linewidth}
\center{\includegraphics[width=0.85\linewidth]{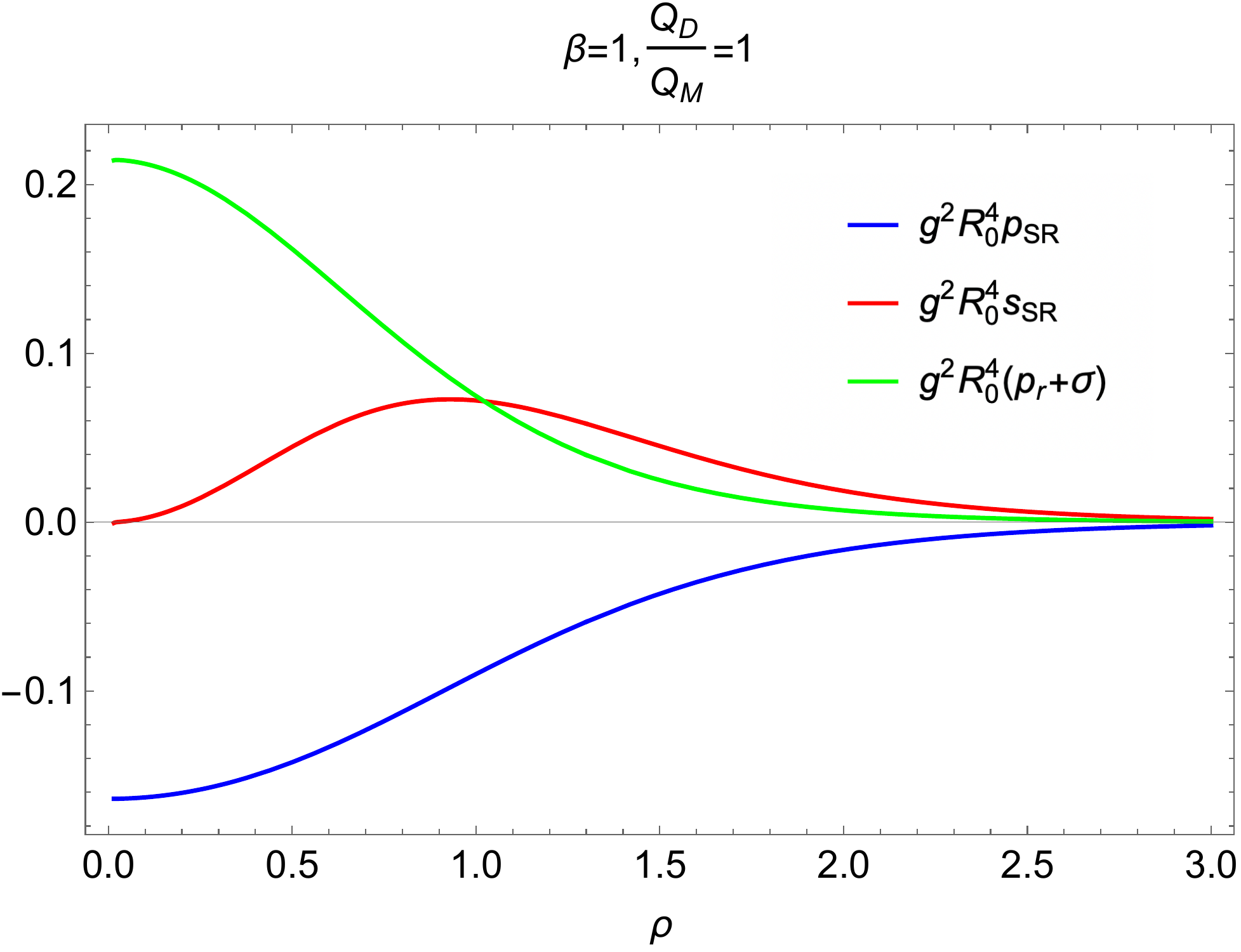}} \\
\end{minipage}
\hfill
\begin{minipage}[h]{0.495\linewidth}
\center{\includegraphics[width=0.85\linewidth]{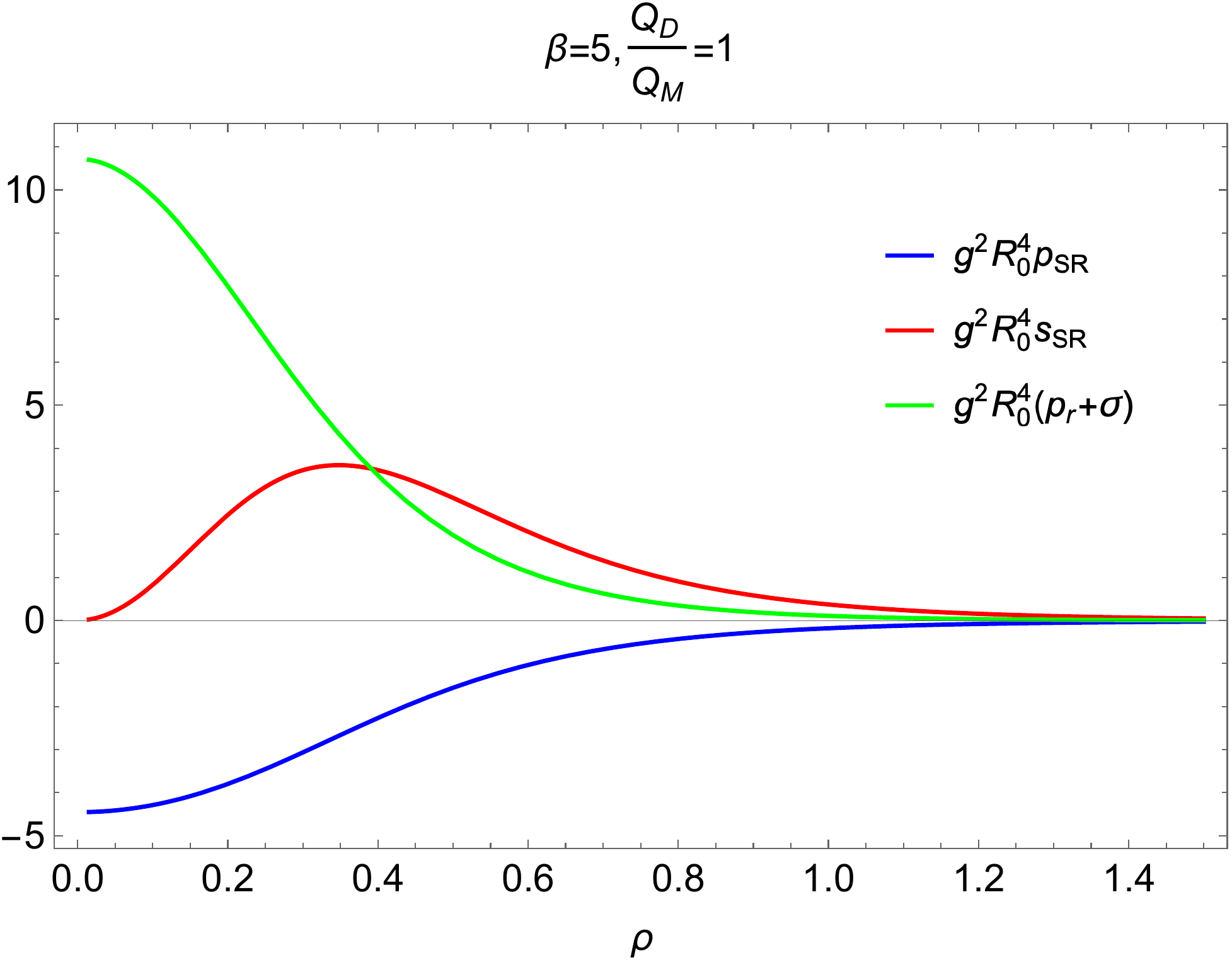}} \\
\end{minipage}
\caption{The short-range mechanical properties of dyons as a functions of $\rho=r/R_0=gvr$ from Eq.~\eqref{EMTDyonSR} for varied value of $\beta$ and charge ratio $\dfrac{Q_D}{Q_M}$. Note, that the total normal force is defined as in Eq.~\eqref{stabilityJ}.}
\label{DmechpropF}
\end{figure}

We will exclude the long-range contribution given in Eq.~\eqref{coulombful} from the EMT of dyon in Eq.~\eqref{T00d} in the same way as we did it for the monopole case using Eq.~\eqref{TSR}. After simple algebraic calculations we obtain  the following expression for the  short-range part of the EMT of the dyon
\begin{equation}\label{EMTDyonSR}
\begin{split}
T_{00}^{\text{SR}}(\rho)&=T_{00}^{\text{M,SR}}(\rho)+\dfrac{1}{R_0^4g^2}\left(\dfrac{1}{2}\tilde{J}'^2\left(1-h^2\right)+\dfrac{\tilde{J}^2F^2}{\rho^2}\right),\\
p^{\text{SR}}(\rho)&=p^{\text{M,SR}}(\rho)+\dfrac{1}{R_0^4g^2}\left(\dfrac{1}{6}\tilde{J}'^2\left(1-h^2\right)+\dfrac{1}{3}\dfrac{\tilde{J}^2F^2}{\rho^2}\right),\\
s^{\text{SR}}(\rho)&=s^{\text{M,SR}}(\rho)-\dfrac{1}{R_0^4g^2}\left(\tilde{J}'^2\left(1-h^2\right)-\dfrac{\tilde{J}^2F^2}{\rho^2}\right).
\end{split}
\end{equation}
Here $T_{00}^{\text{M,SR}}$, $p^{\text{M,SR}}$ and $s^{\text{M,SR}}$ correspond to the short-range part of the EMT of the monopole in Eq.~\eqref{tij_intris}. The final short-range distributions of mechanical properties  are shown in Fig.~\ref{DmechpropF}. These distributions have the same behaviour as in the monopole case.  

The  equilibrium equation of the long- and short-range parts of the EMT analogously  to Eq.~\eqref{equilibriumC} is given as
\be\label{equilCDyon}
\begin{split}
\dfrac{d}{dr}\left(\dfrac{2}{3}s^C(r)+p^C(r)\right)+2\dfrac{s^C(r)}{r}=-\dfrac{Q_M(r)\rho_M(r)}{4\pi r^2}-\dfrac{\tilde{Q}_D(r)\rho_D(r)}{4\pi r^2},\\
\dfrac{d}{dr}\left(\dfrac{2}{3}s^{SR}(r)+p^{SR}(r)\right)+2\dfrac{s^{SR}(r)}{r}=\dfrac{Q_M(r)\rho_M(r)}{4\pi r^2}+\dfrac{\tilde{Q}_D(r)\rho_D(r)}{4\pi r^2},
\end{split}
\ee
where $Q_M(r)$ and $\tilde{Q}_D(r)$ are magnetic and electric charges contained in the sphere of radius $r$:
\be
\bga
Q_M(r)=\smashoperator{\int\limits_{|\vec{x}|<r}}d^3x\rho_{M}(\vec{x})=\dfrac{4\pi}{g}Q(r),\hspace{15pt}\tilde{Q}_D(r)=\smashoperator{\int\limits_{|\vec{x}|<r}}d^3x\rho_{D}(\vec{x})=\dfrac{4\pi}{g}\tilde{Q}(r).
\ega
\ee 
The first equilibrium  equation describes the balance between two types of "Coulomb forces", namely magnetic and electric ones, pulling the dyon inwards to  the center and the "Coulomb stress" pushing the dyon outwards. The second  equation describes the balance between the two types of the repulsive "Coulomb forces" pushing the dyon outward and the "short-range stress" pulling the dyon inwards to the center. Same as for the monopole case it is not possible to define the short-range part of the EMT using the ambiguity of the $\mathcal{F}_{\mu\nu}$ in such a way that the long- and short-range parts of the  EMT are conserved separately unless the electric and magnetic charge densities both vanish. 
 \begin{table}[h]
                           \begin{center}
                           \begin{tabular}{c|c|c|c|c|c|c|c|c|c|c}
                           $\beta$ & $\frac{Q_D}{Q_M}$  &$M_M \frac{R_0g^2}{4\pi}$ &$M \frac{R_0g^2}{4\pi}$&                $M_C\frac{R_0g^2}{4\pi}$ &$M_{SR}\frac{R_0g^2}{4\pi}$  & $-\dfrac{15g^4}{64\pi^2}D$ & $\dfrac{\langle r^2\rangle_{E}}{R_0^2} $& $\dfrac{\langle r^2\rangle_{M}}{R_0^2}$  & $\dfrac{\langle r^2\rangle_D}{R_0^2}$  & $\dfrac{\langle r^2\rangle_{\text{mech}}}{R_0^2}$  \\ \hline
                             $\varepsilon$ & 0.5&0.998 &1.109  & 0.168 & 0.941&   $\infty$ &   $\infty$& $\infty$ &$\infty$&   $\infty$\\ 
                              $\varepsilon$ & 1& 1.050&1.400  &0.207 &1.193  &   $\infty$ &   $\infty$& $\infty$ &$\infty$&   $\infty$\\  \hline        
                   0.1& 0.5&   1.043& 1.164&  0.21&  0.953& 10.826 &    13.863&   23& 21.447&   24.000 \\              
                   0.1& 1&  1.099&  1.484& 0.263&  1.215&  20.832 &      18.373&  29.247&  28.838&    33.253\\  \hline  
                 0.5& 0.5&   1.142&  1.281& 0.304&  0.977& 1.33  &    4.327&  4.803 &   5.624&    3.883\\
                 0.5& 1&  1.2& 1.654& 0.394&  1.26&  2.225 &   6.136&   6.08& 7.732&   4.972 \\ \hline
                 1& 0.5 & 1.220&  1.371&0.386& 0.985& 0.529 &   2.909&   2.416&   3.504&  1.630  \\
                 1& 1 & 1.28&  1.738&0.477& 1.262& 0.698 &    4.467& 2.954&   5.288&   1.200 \\\hline
                 5&  0.5&  1.44& 1.604& 0.839&  0.375& 0.49 &  2.139&   0.4& 2.201& 0.42  \\
                 5& 1&  1.492&1.825& 0.909&  0.915& 0.405 &    3.299&   2.163&   3.489& 0.365
                           \end{tabular}
                           \caption{Numerical value of $D$-terms, radii  and different contributions to the mass for various values of $\beta$ and the charge ratio $\frac{Q_D}{Q_M}$, where $\varepsilon\ll 1$. The estimation of the accuracy of our numerical method is  about 95 percent  for $0\leq\beta\leq1$ and about 85 percent for $\beta> 1$.}
                           \label{mass-dtermDyon}
                           \end{center}
                          \end{table}  
                          
 In table  \ref{mass-dtermDyon} the full mass of dyon is denoted as $M$, the monopole contribution to the full dyon mass as $M_M$, the $M_C$ and $M_{SR}$ are the long- and short-range contributions to the full mass, correspondingly, the  $D$-terms are computed with the help of the short-range part of the shear  force distribution in Eq.~\eqref{EMTDyonSR}. The following quantities can be also  found in the table: the energy mean square radius $\langle r^2\rangle_{E}$, the magnetic  and electric charge mean square radii $\langle r^2\rangle_{M}$ and $\langle r^2\rangle_D$, and the mechanical radius $\langle r^2\rangle_{\text{mech}}$.                 
It is remarkable that the  $D$-term is growing with the parameter $\frac{Q_D}{Q_M}$ and decreasing with the parameter $\beta$. As it was already mentioned the main contribution to the dyon mass is given by the monopole part of the dyon, moreover the same as it was for the monopole: for small values of $\beta$ the main contribution to the mass comes from the short range region, while for the value $\beta\gtrsim 5$ from the long-range part.  The following ratios of the radii can be noticed from the table 
\be
\begin{split}
&\dfrac{\langle r^2\rangle_{M}}{\langle r^2\rangle _{D}}>1, \hspace{10pt} \text{for small $\beta$ and every $\dfrac{Q_D}{Q_M}$},\\
&\dfrac{\langle r^2\rangle_{M}}{\langle r^2\rangle_{D}}<1, \hspace{10pt} \text{for $\beta\gtrsim 0.1$ and every $\dfrac{Q_D}{Q_M}$},\\
&\dfrac{\langle r^2\rangle_{M}}{\langle r^2\rangle _{E}}>1, \hspace{10pt} \text{for $\beta<1$ and every $\dfrac{Q_D}{Q_M}$},\\
&\dfrac{\langle r^2\rangle_{M}}{\langle r^2\rangle_{E}}<1, \hspace{10pt} \text{for $\beta\geq 1 $ and every $\dfrac{Q_D}{Q_M}$ },\\
&\dfrac{\langle r^2\rangle_D}{\langle r^2\rangle_E}>1, \hspace{10pt} \text{for every $\beta$ and every  $\dfrac{Q_D}{Q_M}$}.
\end{split}
\ee
                          
 \section{Summary and conclusions}
\label{summary}
In this work the EMTs of the 't Hooft-Polyakov monopole and the Julia-Zee dyon are studied. These EMTs  contain long-range contributions which present analogies to the "Coulomb interaction".  The local stability condition containing the pressure and shear force distributions  is  violated for both cases, the  monopole and the dyon. Thereby, the applicability of the condition in  the presence of long-range contributions is questioned. Moreover, such important quantities  which give information about mechanical properties of a system as the $D$-term, energy and charge mean square radii of  the monopole and the dyon can not be computed due to the presence of the long-range interaction. 

To shed more light on the mechanical properties of the monopole and the dyon and on the local stability condition we exclude the long-range contribution from the EMT of the monopole and the dyon. The difficulty of such calculation is that this contribution  can not be uniquely defined. We suggest the modified 't Hooft  definition of the abelian  field strength tensor for this purpose. 
Dealing with the separate long- and short-range parts of the EMTs we obtained  the equilibrium equations which couple the  pressure, shear force distributions and the external force acting on the system. In this line we  have also modified the local mechanical stability condition for systems in the presence of external forces. According to this condition the short-range part of the 't Hooft-Polyakov monopole as well as of the Julia-Zee dyon is stable for every choice of the parameter $\beta$. 

Further, in this paper numerous figures describing mechanical properties of the 't Hooft-Polyakov monopole and the Julia-Zee dyon can be found as well as tables with varied contributions to the masses, $D$-terms and mean square energy, magnetic and electric charge radii.  

\acknowledgements
This paper is an extension of my master thesis, which was supervised by Maxim Polyakov in 2021, who passed away in August 2021. I dedicate this paper to his memory.

I would like to thank  H. Alharazin, J. Gegelia, S. Korenblit,  P. Schweitzer and Ya. Shnir for encouraging, useful comments and numerous discussions.   
This work was supported in part by DFG and NSFC through funds provided to the Sino-German CRC 110 “Symmetries and the Emergence of Structure in QCD” (NSFC Grant No. 11621131001, DFG Project-ID 196253076 - TRR 110).

\newpage
\appendix
\section{Asymptotic behaviours}
We specify here the asymptotic behaviours of EMT which  was not directly given in the main body of the paper.
\subsection{For monopole part}
The function $Q(\rho)$ has the following asymptotic behaviour 
\begin{equation}\label{Qbeh}
\begin{split}
Q(\rho)&\underset{\rho\to \infty}\simeq 1-C_h\dfrac{e^{-\beta \rho}}{\rho}\left(1+O\left(\dfrac{1}{\rho}\right)\right)-\dfrac{2C_F^2}{\beta^2-4}\dfrac{e^{-2\rho}}{\rho^2}\left(1+O\left(\dfrac{1}{\rho}\right)\right),\\
Q(\rho)&\underset{\rho\to 0}\simeq \rho^3\left(-2 a b + b^3\right)+\dfrac{\rho^5}{20} \left(-48 a^2 b + 2 a b (32 b^2 + \beta^2) - 
   b^3 (4 + 3 \beta^2)\right)+O(\rho^7).\\
   \end{split}
\end{equation}
The asymptotic behaviour of the magnetic charge density is 
\be\label{asymptotCharge}
\begin{split}
\rho_{M}(\rho)&\underset{\rho\to 0}\simeq \dfrac{1}{g R_0^3}\left[3b^3-6ab+\rho^2\left(-12 a^2 b - b^3 + 16 a b^3 + \dfrac{1}{2} a b \beta^2 - \dfrac{3 b^3 \beta^2}{4}\right)+O(\rho^4)\right],\\
\rho_{M}(\rho)&\underset{\rho\to\infty}\simeq \dfrac{1}{g R_0^3}\left[C_h\beta\dfrac{e^{-\beta \rho}}{\rho^3}\left(1+O\left(\dfrac{1}{\rho}\right)\right)+\dfrac{4C_F^2}{\beta^2-4}\dfrac{e^{-2\rho}}{\rho^4}\left(1+O\left(\dfrac{1}{\rho}\right)\right)\right].\\
\end{split}
\ee

The "Coulomb EMT" behaves asymptotically as
\begin{equation}\label{asymptotC}
\begin{gathered}
T_{00}^{C}(\rho)\underset{\rho\to \infty}\simeq\dfrac{1}{R_0^4g^2}\Bigg[\dfrac{1}{2\rho^4}-C_h\dfrac{e^{-\beta \rho}}{\rho^5}\left(1+O\left(\dfrac{1}{\rho}\right)\right)+\dfrac{C_h^2}{2}\dfrac{e^{-2\beta \rho}}{\rho^6}\left(1+O\left(\dfrac{1}{\rho}\right)\right)-\dfrac{2C_F^2}{\beta^2-4}\dfrac{e^{-2\rho}}{\rho^6}\left(1+O\left(\dfrac{1}{\rho}\right)\right) \Bigg],\\
p^{C}(\rho)\underset{\rho\to \infty}\simeq\dfrac{1}{R_0^4g^2}\Bigg[\dfrac{1}{6\rho^4}-\dfrac{C_h}{3}\dfrac{e^{-\beta \rho}}{\rho^5}\left(1+O\left(\dfrac{1}{\rho}\right)\right)+\dfrac{C_h^2}{6}\dfrac{e^{-2\beta \rho}}{\rho^6}\left(1+O\left(\dfrac{1}{\rho}\right)\right)-\dfrac{2C_F^2}{3(\beta^2-4)}\dfrac{e^{-2\rho}}{\rho^6}\left(1+O\left(\dfrac{1}{\rho}\right)\right)\Bigg],\\
s^{C}(\rho)\underset{\rho\to \infty}\simeq\dfrac{1}{R_0^4g^2}\Bigg[-\dfrac{1}{\rho^4}+2C_h\dfrac{e^{-\beta \rho}}{\rho^5}\left(1+O\left(\dfrac{1}{\rho}\right)\right)-C_h^2\dfrac{e^{-2\beta \rho}}{\rho^6}\left(1+O\left(\dfrac{1}{\rho}\right)\right)+\dfrac{4C_F^2}{\beta^2-4}\dfrac{e^{-2\rho}}{\rho^6}\left(1+O\left(\dfrac{1}{\rho}\right)\right)\Bigg],\\
T_{00}^{C}(\rho)\underset{\rho\to 0}\simeq\dfrac{1}{R_0^4g^2}\Bigg[\dfrac{\rho^2}{2}\left(-2 a b + 
  b^3\right)^2+\dfrac{\rho^4} {20} b \left(-2 a + b^2\right) \left(-48 a^2 b + 
   2 a b (32 b^2 + \beta^2) - b^3 (4 + 3 \beta^2)\right)+O(\rho^6)\Bigg],\\
p^{C}(\rho)\underset{\rho\to 0}\simeq\dfrac{1}{R_0^4g^2}\Bigg[\dfrac{\rho^2}{6}\left(-2 a b + 
  b^3\right)^2 +\dfrac{\rho^4} {60} b \left(-2 a + b^2\right) \left(-48 a^2 b + 
   2 a b (32 b^2 + \beta^2) - b^3 (4 + 3 \beta^2)\right)+O(\rho^6)\Bigg],\\
s^{C}(\rho)\underset{\rho\to 0}\simeq -\dfrac{1}{R_0^4g^2}\Bigg[\rho^2\left(-2 a b + 
  b^3\right)^2 +\dfrac{\rho^4} {10} b \left(-2 a + b^2\right) \left(-48 a^2 b + 
   2 a b (32 b^2 + \beta^2) - b^3 (4 + 3 \beta^2)\right)+O(\rho^6)\Bigg].
   \end{gathered}
\end{equation}

The asymptotic behaviour of the short-range part of the EMT is
\begin{equation}\label{asymptSR}
\begin{gathered}
T_{00}^{SR}(\rho)\underset{\rho\to \infty}\simeq\dfrac{1}{R_0^4g^2}\Bigg[C_h\dfrac{e^{-\beta\rho}}{\rho^5}\left(1+O\left(\dfrac{1}{\rho}\right)\right)+C_h^2\beta^2 \dfrac{e^{-2\beta\rho}}{\rho^2}\left(1+O\left(\dfrac{1}{\rho}\right)\right)+2C_F^2\dfrac{e^{-2\rho}}{\rho^2}\left(1+O\left(\dfrac{1}{\rho}\right)\right)\Bigg],\\
p^{SR}(\rho)\underset{\rho\to \infty} \simeq\dfrac{1}{R_0^4g^2}\Bigg[\dfrac{C_h}{3}\dfrac{e^{-\beta \rho}}{\rho^5}\left(1+O\left(\dfrac{1}{\rho}\right)\right)-\dfrac{2}{3}C_h^2\beta^2 \dfrac{e^{-2\beta\rho}}{\rho^2}\left(1+O\left(\dfrac{1}{\rho}\right)\right)-\dfrac{2C_F^2}{3}\dfrac{e^{-2\rho}}{\rho^4}\left(1+O\left(\dfrac{1}{\rho}\right)\right)\Bigg],\\
s^{SR}(\rho)\underset{\rho\to \infty}\simeq  \dfrac{1}{R_0^4g^2}\Bigg[-2C_h\dfrac{e^{-\beta \rho}}{\rho^5}\left(1+O\left(\dfrac{1}{\rho}\right)\right)+C_h^2\beta^2 \dfrac{e^{-2\beta\rho}}{\rho^2}\left(1+O\left(\dfrac{1}{\rho}\right)\right)+C_F^2\dfrac{e^{-2\rho}}{\rho^4}\left(1+O\left(\dfrac{1}{\rho}\right)\right)\Bigg],\\
T_{00}^{SR}(\rho)\underset{\rho\to 0}\simeq \dfrac{1}{R_0^4g^2}\Bigg[\left(6a^2+\dfrac{3}{2}b^2+\dfrac{\beta^2}{8}\right)+\left(8a^3+6ab^2-2a^2b^2+2ab^4-\dfrac{b^6}{2}-\dfrac{\beta^2b^2}{2}\right)\rho^2+O(\rho^4)\Bigg],\\
p^{SR}(\rho)\underset{\rho\to 0}\simeq \dfrac{1}{R_0^4g^2}\Bigg[\left(2a^2-\dfrac{b^2}{2}-\dfrac{\beta^2}{8}\right)+\left(8a^3-2ab^2-2a^2b^2+2ab^4-\dfrac{b^6}{2}+\beta^2b^2\right)\dfrac{\rho^2}{3}+O(\rho^4)\Bigg],\\
s^{SR}(\rho)\underset{\rho\to 0}\simeq \dfrac{1}{R_0^4g^2}\left[\left(-8a^3+2ab^2+20a^2b^2-20ab^4+5b^6-\beta^2b\right)\dfrac{\rho^2}{5}+O(\rho^4)\right].
\end{gathered}
\end{equation}

\subsection{For dyon part}
\label{DyonApp}
The function $\tilde{Q}(\rho)$ has the following asymptotic behaviour 
\begin{equation}
\begin{split} 
\tilde{Q}(\rho)& \underset{\rho\to \infty}\simeq  g Q_D-g C_h Q_D\dfrac{e^{-\beta \rho}}{\rho}\left(1+O\left(\dfrac{1}{\rho}\right)\right)-\dfrac{C C_F^2}{\sqrt{1 - C^2}}e^{-2\sqrt{1 - C^2} \rho}\left(1+O\left(\dfrac{1}{\rho}\right)\right),\\
\tilde{Q}(\rho)&\underset{\rho\to 0}\simeq \rho^3 b c+\rho^5 \left(\dfrac{6 a b c}{5} + \dfrac{c}{20} (8 a b - b \beta^2) \right)+O(\rho^7).
\end{split}
\end{equation}
The asymptotic behaviour of the electric charge density is 
\begin{equation}
\begin{split} 
\rho_D(\rho)& \underset{\rho\to \infty}\simeq  \dfrac{1}{gR_0^3}\left(gQ_D C_h\beta\dfrac{e^{-\beta \rho}}{\rho^3}\left(1+O\left(\dfrac{1}{\rho}\right)\right)+2C C_F^2\dfrac{e^{-2\sqrt{1 - C^2} \rho}}{\rho^2}\left(1+O\left(\dfrac{1}{\rho}\right)\right)\right),\\
\rho_D(\rho)&\underset{\rho\to 0}\simeq\dfrac{1}{gR_0^3}\left(3 b c+\rho^2\left(6 a b c + \dfrac{c}{4} (8 a b - b \beta^2) \right)\right)+O(\rho^4).
\end{split}
\end{equation}

The asymptotic behaviour of the full EMT of the dyon near the origin is
\be\label{emtD0}
\begin{split}
T_{00}(\rho)&\underset{\rho\to 0}\simeq\dfrac{1}{R_0^4g^2}\left[6 a^2 + \dfrac{3 b^2}{2} + \dfrac{\beta^2}{8} + \dfrac{3 c^2}{2}+\rho^2\left( 
8 a^3 + 6 a b^2 - \dfrac{b^2 \beta^2}{2} + 2 a c^2\right)+O(\rho^4)\right], \\
p(\rho)&\underset{\rho\to 0}\simeq \dfrac{1}{R_0^4g^2}\left[2 a^2 - \dfrac{b^2}{2} - \dfrac{\beta^2}{8} + \dfrac{c^2}{2}+ \dfrac{\rho^2}{3}\left(8 a^3 - 2 a b^2 + b^2 \beta^2 + 2 a c^2\right)+O(\rho^4)\right],\\
s(\rho)& \underset{\rho\to 0}\simeq \dfrac{1}{R_0^4g^2}\left[\dfrac{\rho^2}{5}\left(-8 a^3+2 a b^2- b^2 \beta^2 - 2 a c^2\right)+O(\rho^4)\right].
\end{split}
\ee 

The asymptotic behaviour of the electric part of the full EMT for the dyon:
\be
\begin{split}
T_{00}^E(\rho)&\underset{\rho\to \infty}\simeq \dfrac{g^2Q_D^2}{2}\dfrac{1}{\rho^4}+C^2C_F^2\dfrac{e^{-2\sqrt{1-C^2}\rho}}{\rho^2}\left(1+O\left(\dfrac{1}{\rho}\right)\right), \\
p^E(\rho)&\underset{\rho\to \infty}\simeq \dfrac{g^2Q_D^2}{6}\dfrac{1}{\rho^4}+\dfrac{C^2C_F^2}{3}\dfrac{e^{-2\sqrt{1-C^2}\rho}}{\rho^2}\left(1+O\left(\dfrac{1}{\rho}\right)\right),\\
s^E(\rho)&\underset{\rho\to \infty}\simeq -g^2Q_D^2\dfrac{1}{\rho^4}+C^2C_F^2\dfrac{e^{-2\sqrt{1-C^2}\rho}}{\rho^2}\left(1+O\left(\dfrac{1}{\rho}\right)\right),\\
T_{00}^E(\rho)&\underset{\rho\to 0}\simeq \dfrac{3 c^2}{2} + 4 a c^2 \rho^2+O(\rho^4), \\
p^E(\rho)&\underset{\rho\to 0}\simeq  \dfrac{c^2}{2} + \dfrac{4}{3} a c^2 \rho^2+O(\rho^4),\\
s^E(\rho)&\underset{\rho\to 0}\simeq\dfrac{2}{5} a c^2 \rho^2+O(\rho^4).
\end{split}
\ee 

The asymptotic behaviour of the long-range part of the EMT is 
\be\label{CoulombAsympD}
\bga
T_{00}^{C}(\rho)\underset{\rho\to \infty}\simeq \dfrac{1}{R_0^4g^2}\Bigg[\dfrac{1}{2}(1+g^2 Q^2_D)\dfrac{1}{\rho^4}-C_h(1+g^2Q_D^2)\dfrac{e^{-\beta \rho}}{\rho^5}\left(1+O\left(\dfrac{1}{\rho}\right)\right)-\\
-\dfrac{C C_F^2 gQ_D}{\sqrt{1 - C^2}}\dfrac{e^{-2\sqrt{1 - C^2}\rho}}{\rho^4}\left(1+O\left(\dfrac{1}{\rho}\right)\right) +\dfrac{C_h^2}{2}(1+g^2Q_D^2)\dfrac{e^{-2\beta \rho}}{\rho^6}\left(1+O\left(\dfrac{1}{\rho}\right)\right)\Bigg],\\
p^{C}(\rho)\underset{\rho\to \infty}\simeq \dfrac{1}{R_0^4g^2}\Bigg[\dfrac{1}{6}(1+g^2 Q^2_D)\dfrac{1}{\rho^4}-\dfrac{C_h}{3}(1+g^2Q^2_D)\dfrac{e^{-\beta \rho}}{\rho^5}\left(1+O\left(\dfrac{1}{\rho}\right)\right)-\\
-\dfrac{C C_F^2 gQ_D}{3\sqrt{1 - C^2}}\dfrac{e^{-2\sqrt{1 - C^2}\rho}}{\rho^4}\left(1+O\left(\dfrac{1}{\rho}\right)\right) +\dfrac{C_h^2}{2}(1+g^2Q_D^2)\dfrac{e^{-2\beta \rho}}{\rho^6}\left(1+O\left(\dfrac{1}{\rho}\right)\right)\Bigg],\\
s^{C}(\rho)\underset{\rho\to \infty}\simeq\dfrac{1}{R_0^4g^2}\Bigg[ -(1+g^2 Q^2_D)\dfrac{1}{\rho^4}+2C_h(1+g^2Q^2_D)\dfrac{e^{-\beta \rho}}{\rho^5}\left(1+O\left(\dfrac{1}{\rho}\right)\right)+\\
+\dfrac{2C C_F^2 gQ_D}{\sqrt{1 - C^2}}\dfrac{e^{-2\sqrt{1 - C^2}\rho}}{\rho^4}\left(1+O\left(\dfrac{1}{\rho}\right)\right) -C_h^2(1+g^2Q_D^2)\dfrac{e^{-2\beta \rho}}{\rho^6}\left(1+O\left(\dfrac{1}{\rho}\right)\right)\Bigg], \\
T_{00}^{C}(\rho)\underset{\rho\to 0}\simeq\dfrac{1}{R_0^4g^2}\left[ \rho^4 \dfrac{1 }{2}  b^2 ((-2 a + b^2)^2 + c^2)+O(\rho^6)\right],\\
p^{C}(\rho)\underset{\rho\to 0}\simeq\dfrac{1}{R_0^4g^2}\left[ \rho^4 \dfrac{1}{6}  b^2 ((-2 a + b^2)^2 + c^2)+O(\rho^6)\right],\\
s^{C}(\rho)\underset{\rho\to 0}\simeq\dfrac{1}{R_0^4g^2}\left[-\rho^4 b^2 ((-2 a + b^2)^2 + c^2)+O(\rho^6)\right].
\ega
\ee

The asymptotic behaviour of the short-range part of the  EMT of the dyon is as follows:
\be\label{SRDyonAsym}
\bga
T_{00}^{SR}(\rho)\underset{\rho\to \infty}\simeq \dfrac{1}{R_0^4g^2}\Bigg[2C_F^2\dfrac{e^{-2\sqrt{1-C^2}\rho}}{\rho^2}\left(1+O\left(\dfrac{1}{\rho}\right)\right)+\beta^2C_F^2\dfrac{e^{-2\beta\rho}}{\rho^2}\left(1+O\left(\dfrac{1}{\rho}\right)\right)+C_h(1+g^2Q_D^2)\dfrac{e^{-\beta \rho}}{\rho^5}\left(1+O\left(\dfrac{1}{\rho}\right)\right)\Bigg], \\
p^{SR}(\rho)\underset{\rho\to \infty}\simeq\dfrac{1}{R_0^4g^2}\Bigg[ -\dfrac{2}{3}\beta^2C_F^2\dfrac{e^{-2\beta\rho}}{\rho^2}\left(1+O\left(\dfrac{1}{\rho}\right)\right)+O\left(\dfrac{e^{-2\sqrt{1-C^2}\rho}}{\rho^3}\right)+\dfrac{C_h}{3}(1+g^2Q^2_D)\dfrac{e^{-\beta \rho}}{\rho^5}\left(1+O\left(\dfrac{1}{\rho}\right)\right)\Bigg],\\
s^{SR}(\rho)\underset{\rho\to \infty}\simeq \dfrac{1}{R_0^4g^2}\Bigg[\beta^2C_F^2\dfrac{e^{-2\beta\rho}}{\rho^2}\left(1+O\left(\dfrac{1}{\rho}\right)\right)+O\left(\dfrac{e^{-2\sqrt{1-C^2}\rho}}{\rho^3}\right)-2C_h(1+g^2Q^2_D)\dfrac{e^{-\beta \rho}}{\rho^5}\left(1+O\left(\dfrac{1}{\rho}\right)\right)\Bigg],\\
T_{00}^{SR}(\rho)\underset{\rho\to 0}\simeq\dfrac{1}{R_0^4g^2}\left(6 a^2 + \dfrac{3 b^2}{2} + \dfrac{\beta^2}{8} + \dfrac{3 c^2}{2}+\rho^2\left[ 
8 a^3 + 6 a b^2 - \dfrac{b^2 \beta^2}{2} + 2 a c^2\right]+O(\rho^4)\right), \\
p(\rho)^{SR}\underset{\rho\to 0}\simeq \dfrac{1}{R_0^4g^2}\left(2 a^2 - \dfrac{b^2}{2} - \dfrac{\beta^2}{8} + \dfrac{c^2}{2}+ \dfrac{\rho^2}{3}\left[8 a^3 - 2 a b^2 + b^2 \beta^2 + 2 a c^2\right]+O(\rho^4)\right),\\
s(\rho)^{SR}\underset{\rho\to 0}\simeq \dfrac{1}{R_0^4g^2}\left(\dfrac{\rho^2}{5}\left[-8 a^3+2 a b^2- b^2 \beta^2 - 2 a c^2\right]+O(\rho^4)\right).
\ega
\ee

\section{Divergence of radii}
\label{divR}
As it follows from the Eqs.~\eqref{asymptotCharge} and \eqref{asymptSR} the main contribution to the magnetic charge density and  to the short-range part of the energy density, respectively,  is provided by the asymptotic behaviour at large-distances. Then the corresponding mean square radii  for the small values of $\beta$ and for the large enough $R$  diverge as    
\be
\bga
\label{RMdiv}
\dfrac{\langle r_M^2\rangle}{R_0^2}\simeq\frac{\int\limits_{R}^{\infty} d\rho \rho^4\left(C_h\beta\dfrac{e^{-\beta\rho}}{\rho^3}\right)}{\int\limits_{R}^{\infty} d\rho \rho^2\left(C_h\beta\dfrac{e^{-\beta\rho}}{\rho^3}\right)}\sim\frac{\int\limits_{R}^{1/\beta} d\rho \rho}{\int \limits_{R}^{1/\beta}d\rho \dfrac{1}{\rho}}\sim \dfrac{1}{\beta^2\ln\left(\frac{1}{\beta}\right)},
\ega
\ee
\be
\bga\label{REdiv}
\dfrac{\langle r_E^2\rangle}{R_0^2}\simeq\frac{\int\limits_{R}^{\infty} d\rho \rho^4\left(C_h\dfrac{e^{-\beta\rho}}{\rho^5}+C_h^2\beta^2\dfrac{e^{-2\beta\rho}}{\rho^2}\right)}{\int\limits_{R}^{\infty} d\rho \rho^2\left(C_h\dfrac{e^{-\beta\rho}}{\rho^5}+C_h^2\beta^2\dfrac{e^{-2\beta\rho}}{\rho^2}\right)}\sim\frac{\int\limits_{R}^{1/\beta} d\rho \left( \dfrac{1}{\rho}+\beta^2\rho^2\right)}{\int \limits_{R}^{1/\beta}d\rho\left( \dfrac{1}{\rho^3}+\beta^2\right)}\sim \dfrac{1}{\beta^2}-\dfrac{\ln\beta}{\beta}.
\ega
\ee
According to the modified definition of the mechanical radius in Eq.~\eqref{r^2mech2}, to the external force in equilibrium equation \eqref{equilibriumC} for the short-range and to the asymptotic behaviour in Eqs.~\eqref{Qbeh} and \eqref{asymptSR} the mean square mechanical radius of the monopole for small values of $\beta$ diverges as follows
\be
\bga\label{Rmechdiv}
\dfrac{\langle r^2\rangle _{\text{mech}}}{R_0^2}\sim\frac{\int\limits_{R}^{1/\beta} d\rho \rho^4\left( \dfrac{\beta}{\rho^4}-\dfrac{1}{\rho^5}\right)}{\int \limits_{R}^{1/\beta}d\rho\rho^2\left( \dfrac{\beta}{\rho^4}-\dfrac{1}{\rho^5}\right)}\sim \dfrac{\ln\beta}{\beta^2}.
\ega
\ee
 
Proceeding analogously to the monopole case, we obtain for the dyon the same divergences as in Eqs.~\eqref{Ddiverge}, \eqref{RMdiv}, \eqref{REdiv} and \eqref{Rmechdiv}. Namely, the $D$-term diverges as in Eq.~\eqref{Ddiverge}, the mean square electric and magnetic charge radii as in Eq.~\eqref{RMdiv}, the mean square energy radius of the short-range part as in Eq.~\eqref{REdiv} and the mean square mechanical radius as in Eq.~\eqref{Rmechdiv}.

\end{document}